\documentclass[twocolumn,9pt,a4paper,fleqn]{article}  

\usepackage[left=0.6in, right=0.6in, top=0.6in, bottom=1in]{geometry} 
\usepackage{amsmath}
 \numberwithin{equation}{section}

\usepackage{setspace}
\RequirePackage{graphicx}
\RequirePackage{newtxtext}       
\RequirePackage{flushend}
\RequirePackage[numbers,sort&compress]{natbib}
\RequirePackage[colorlinks,citecolor=blue,urlcolor=blue,linkcolor=blue]{hyperref}
\usepackage{newtxmath}
\usepackage{doi}
\usepackage{orcidlink}

\newcommand{\bq}{\begin{equation}}
	\newcommand{\eq}{\end{equation}}
\newcommand{\bqn}{\begin{eqnarray}}
	\newcommand{\eqn}{\end{eqnarray}}

\newcommand{\zb}{\ensuremath{\bar{z}}}
\newcommand{\xb}{\ensuremath{\bar{x}}}

\RequirePackage{authblk}

\title{Dynamical Systems Approach to Non-Slow-Roll Inflationary Models}

\author[1]{\bf Sandip Biswas\thanks{\href{mailto:sandipb20@iitk.ac.in}{sandipb20@iitk.ac.in}}  }

\affil[1,3]{{\normalsize \it Department of Physics, Indian Institute of Technology Kanpur, Uttar Pradesh, 208016 India}}
		
\author[2]{ \bf Saddam Hussain\orcidlink{0000-0001-6173-6140}\thanks{\href{mailto:saddamh@zjut.edu.cn}{saddamh@zjut.edu.cn}}}

\affil{\normalsize\it Institute for Theoretical Physics and Cosmology, Zhejiang University of Technology, Hangzhou 310023, China}		
			
\author[3]{\bf Kaushik Bhattacharya \thanks{\href{mailto:kaushikb@iitk.ac.in}{kaushikb@iitk.ac.in}} }

\begin{document}
	
	\makeatletter
	\renewcommand{\@date}{\vspace*{-0.5cm}({\normalsize\itshape\today})}
	\makeatother
\maketitle

\begin{abstract}
In this work, we systematically present a new dynamical systems approach to nonstandard inflationary processes as constant-roll inflation and ultraslow-roll inflation. Using the techniques presented in our work one can in general investigate the attractor nature of the inflationary models in the phase space. We have compactified the phase space coordinates, wherever necessary, and regulated the nonlinear differential equations, constituting the autonomous system of equations defining the dynamical system, at the cost of a new redefined time variable which is a monotonic increasing function of the standard time coordinate. We have shown that in most of the relevant cases the program is executable although the two time coordinates may show different durations of cosmological events. Our methods of analysis differs slightly in different models but we have always emphasized on the nature of the initial conditions leading to stable inflationary phases in different cases. 
We have provided a universal language in terms of which various nonstandard inflationary models can be studied.
\end{abstract}

\section{Introduction}

Inflationary models have been pivotal in explaining the early universe's rapid expansion and resolving several cosmological puzzles, such as the flatness and horizon problems \cite{Kazanas:1980tx, Guth:1980zm, Sato:1980yn, Sato:1981ds, Linde:1981mu, Albrecht:1982wi, Riotto:2002yw}. In inflationary dynamics it is tacitly assumed that the very early universe do evolve in an inflationary phase after the initial singularity for a very brief period of time. In this case a question arises, what are the initial conditions which naturally drive the system towards inflation? As it is very difficult to specify some specific set of initial conditions in the very early universe with pin-point accuracy, the general consensus is to focus on a class of initial conditions which can give rise to inflation. If a wide region, in the set of initial conditions on phase space, leads to inflation, one can be sure that an inflation like phase was there after the initial cosmological singularity. The existence of such a large class of initial conditions shows that inflation is an attractor solution. A preliminary but interesting theory of attractor solution of cold inflation (CI) was presented in Ref.~\cite{Mukhanov:1990me, Urena-Lopez:2007zal,Liddle_1994, Carrasco_2015,Alho:2023pkl}. If a wide range of initial conditions can produce inflation, with all its constraints, then we say the attractor solution shows stability.      

In general the stability issues about any system are best studied in the dynamical systems approach where one recasts all the dynamical equations of the system in the form of nonlinear, first order autonomous system of differential equations. The dynamical systems approach has emerged as a powerful tool in late-time cosmology, particularly for analyzing new classes of cosmological models \cite{Boehmer:2014vea, Boehmer:2011tp, Bahamonde:2017ize, Bouhmadi-Lopez:2016dzw, Stachowski_2016, Coley:2003mj,Rendall:2001it,Dutta:2017wfd,blackmore2011nonlinear,elias2006critical,Chatterjee:2021ijw,Hussain:2022dhp,Hussain:2023kwk,Hussain:2024qrd,Bhattacharya:2022wzu, Copeland:2006wr,Saddam:2024xie, Odintsov_2017_01}. In inflationary cosmology, one rarely use the full potential of the dynamical systems approach as we do not expect the system to have any stable fixed points during the inflationary regime. Moreover, a successful period of inflation ends in reheating the universe and the physics of reheating \cite{Kofman:1994rk, Kofman:1997yn, Allahverdi:2010xz, Bassett:2005xm} is different from the physics of slow-roll inflation. Consequently a dynamical system which models a slow-roll process will be unable to yield meaningful information near the end of inflation unless the equations are modified so that the reheating process is also included in the dynamical equations. It is practically impossible to describe the initiation of inflation and the reheating phase with the same set of dynamical equations and mostly the dynamical systems approach concentrates on the initial development of an inflationary system and the analysis can be extended approximately up to the graceful exit phase.  In warm inflation (WI) one does not have a separate reheating phase and we expect that a dynamical systems approach in WI is more useful. A dynamical systems approach has some advantages, it is a very general method which has the potential to unravel the dynamical properties of any system unambiguously. If the dynamical systems approach can be used in conjunction with a wide class of initial conditions then we get a set of trajectories in the phase space showing the qualitative behavior of the system. Except giving an overall, qualitative phase space picture of inflationary dynamics one can also  
analyze any inflationary  scenario in detail using the methods of dynamical systems. In this paper we have tried to posit a uniform dynamical systems viewpoint using which one can study any inflationary process. 

Except cold inflationary paradigm we also have a radically different inflationary paradigm: the paradigm of warm inflation (WI) \cite{Berera:1995ie, Kamali:2023lzq, Berera:2008ar, Bastero-Gil:2016qru, Berghaus:2019whh, Berera:1995wh, Berera:1996nv, Bastero_Gil_2013, Berera_2023, Berera:1999ws, Bastero-Gil:2009sdq, PhysRevD.62.083517, PhysRevD.69.083525, Moss:2011qc, Moss:2007cv, Chen_2008, Gupta:2002kn}. In this paradigm, inflation happens due to the vacuum energy of the inflaton but the inflaton does decay to radiation during inflation and consequently the fluctuations produced by WI are thermal in nature. Warm inflation is more dynamically constrained, than the simple CI models, as in these case the radiation produced from the decay of inflaton can produce a dynamical thermal equilibrium as a result of which the radiation bath can have a temperature $T$. 
The dynamical constraints arises from multiple requirements: 
\begin{itemize}
	
	\item firstly, the requirement of successful inflation when inflaton energy inflates the system and simultaneously decays to radiation.
	
	\item The second requirement of maintaining a stable temperature $T$ during the inflationary process.
	
	\item The third requirement to maintain $T$ greater than the Hubble parameter during WI.
	
\end{itemize}
This last condition is essential for producing thermal fluctuations. Few previous authors have studied the dynamical systems approach and stability issues in WI \cite{deOliveira:1997jt, Moss:2008yb,Li:2018sfs,Das:2023rat,Odintsov:2023lbb,DAgostino:2021vvv,Zhang:2024kcf,Jawad:2017nkq}. 

Closely related to the above two main modern paradigms of inflationary dynamics are some new variants. These variants are all some forms of CI or WI but they differ from the canonical models by the scalar field rolling condition. One of these variants is the constant-roll cold inflation (CRCI) \cite{ Motohashi_2015, Yi:2017mxs, Guerrero:2020lng, Mohammadi:2022tmk, Shokri:2021zqw, Lin:2019fcz, Motohashi_2017, Motohashi_20171, Odintsov_2017, Oikonomou_2017, Gao_2017, Ito_2018, Karam_2018, Cicciarella_2018, Anguelova_2018, Gao_2018, Gao_2018_1, Nojiri_2017, Awad_2018, Odintsov_2017_02}, which is a variant of CI where the slow-roll (SR) condition of the scalar field is changed to the constant-roll condition. In general inflationary dynamics is specified by some rolling conditions and one of these conditions is related to the value of $\ddot{\phi}/(H\dot{\phi})$ where $\phi$ represents the inflaton field and $H$ is the Hubble parameter during inflation. The dot represent derivatives with respect to cosmological time. In the slow-roll regime this ratio tends to zero, whereas in the constant-roll regime this ratio remains a constant. The constant-roll (CR) condition restricts the inflationary system although it is known that it remains an attractor solution near the SR limit. In this work we will verify this claim and show how the attractor solution in CRCI gets modified from the attractor solution in CI.
We will also show that in CRCI there will be some dynamical evolutions where an accelerated expansion phase, like the inflation phase,  goes on inflating and cannot come out of the inflationary process through graceful exit.  

Building on the concepts of (CRCI) and WI, researchers have proposed models of constant-roll warm inflation (CRWI) \cite{Kamali:2019wdh, Biswas:2024oje}. In this variant, one studies warm inflation after imposing the constant-roll condition, i.e. demanding $\ddot{\phi}/(H\dot{\phi})$ to be a constant. CRWI is a highly constrained framework, as it inherits the requirement of dynamical equilibrium in the radiation bath from warm inflation (WI), while the constant-roll condition further restricts the dynamics of the system.
The constant-roll (CR) condition imposes significant restrictions on the CR variants of both cold and warm inflation, permitting only a narrow class of scalar field potentials to sustain these inflationary phases—unlike standard cold or warm inflation, which can be realized with a broad range of potentials \cite{Martin:2005ir}. These constraints render a substantial portion of the phase space inaccessible to the dynamical system. Nevertheless, the attractor behavior of CRWI can still be visualized within the available phase space. In this work, we have presented the dynamics of such systems and have thoroughly analyzed the structure of the phase space for various values of the CR parameters. 

Another variant of the standard SR inflation is the ultraslow-roll (USR) inflation, which happens when $\ddot{\phi}/(H\dot{\phi}) \sim -1$ when $V_{,\phi} \sim 0$, where $V_{,\phi}$ is the derivative of inflaton potential with respect to the inflaton field. In the present paper we have worked in warm USR model \cite{Biswas_2024} where the USR condition is modified. It is well known that USR phase is a transient phase which terminates after a few $e$-folds and after that the SR phase commences. The outcome of our analysis is very interesting in this case. For a particular potential we can show that warm USR inflation does not have a general attractor like solution in the sense that, if the initial conditions do not satisfy the USR condition then USR will not lead to a SR phase. This condition is different from the previous cases where the system was always in a CR regime. CRCI or CRWI are not transient phases, whereas USR  is a transient phase which terminates soon after its onset. In this case one may try to see whether more general initial conditions can naturally give rise to an USR phase.

In this paper, we have independently presented the dynamical analysis for each type of inflation discussed, allowing for a direct comparison of how different inflationary conditions influence the phase space dynamics. A unified framework based on dynamical systems theory has been employed to study the various forms of inflation. We have successfully applied compactified phase space techniques and addressed ill-defined autonomous equations through a redefinition of time. An exception is made in the case of CRCI, where the dynamical phase space is effectively one-dimensional. In such cases, a conventional autonomous system analysis offers limited insight. Therefore, for CRCI, we have adopted an alternative approach using stream plots of the phase space variables in a two-dimensional plane.

In both Cold and Warm Inflation, the background evolution is determined by the flat Friedmann-Lemaitre-Robertson-Walker 
(FLRW) metric described by the line element:  
\begin{equation}
	ds^{2} = dt^{2} - a(t)^{2} d\mathbf{x}^2, \label{FRW_metric}
\end{equation}
where \(t\) is the cosmic time, \(a(t)\) is the scale-factor of the universe, and \(\mathbf{x}\) represents the spatial coordinates \((x,y,z)\).
The fundamental dynamical equations governing a cold inflationary system consist of the Friedmann equations and the Klein–Gordon equation for the inflaton field, formulated in the FLRW background. In the case of warm inflation, the presence of a subdominant yet non-negligible radiation fluid—coupled to the inflaton field—necessitates the inclusion of the radiation fluid's evolution equation as part of the dynamical system. To apply the dynamical systems approach, it is customary to recast the equations into an autonomous system using suitably defined phase space variables. These variables are chosen to ensure that the phase space has the minimal possible dimensionality and that all quantities involved are dimensionless. In our analysis, we reparametrize the cosmic time \( t \) either using a dimensionless time variable \( t' \) in certain cases, or by the number of \( e \)-folds, defined via \( dN = H\,dt \), where \( H \equiv \dot{a}/a \) is the Hubble parameter.

In general, the phase space of a system can be either one-dimensional or higher-dimensional, depending on the number of independent dynamical variables required to construct a closed autonomous system of equations. For an \( n \)-dimensional phase space—corresponding to a cosmological system with \( n \) independent dynamical variables—the first Friedmann equation, which involves the square of the Hubble parameter and various energy densities, always yields a constraint equation of the form:
\begin{eqnarray}
	f(x_1,x_2,\cdots,x_n)=0\,,
	\label{pconstr}
\end{eqnarray}
where $(x_1, x_2, \cdot, \cdot, \cdot)$ are the phase space variables (not to confuse with the spatial coordinates $\mathbf{x}$ in Eq.~(\ref{FRW_metric})) and $f(x_1,x_2,x_3,\cdot\cdot,x_n)$ is an arbitrary function of these variables. The above constraint equation characterizes the topological structure of the phase space for the cosmological system. In such cases, the autonomous system of equations is typically written in the form:
\begin{eqnarray}
	\dot{x}_1 &=& f_1(x_1,x_2,\cdots,x_n)\,,\nonumber\\
	\dot{x}_2 &=& f_2(x_1,x_2,\cdots,x_n)\,,\nonumber\\
	&\vdots&\nonumber\\
	\dot{x}_n &=& f_n(x_1,x_2,\cdots,x_n)\,, \label{gen-dyn-eqs}
\end{eqnarray}
where the overdots specify a derivative with respect to either $t^\prime$ or $N$, which are both dimensionless.  Here the functions $f_j(x_1,x_2,\cdots,x_n)$ (where $j=1,2,\cdots,n$) are in general nonlinear functions of the variables. The form of these functions are obtained from the actual dynamical equations of the system. The critical points (or fixed points) of the above system of equations are obtained for points $x_{c,k} \equiv (x_{1,k}^*, x_{2,k}^*, \cdots, x_{n,k}^*)$ ($k=1,2,\cdots,m$ where $m$ represents the number of critical points) for which
$$\dot{x}_1 = \dot{x}_2  = \cdots = \dot{x}_n=0\,.$$ 
The condition may be satisfied for multiple points in the phase space and all those points ($x_{c,i}$) specify fixed points of the dynamical system.

The stability of the fixed points is determined by linearizing the right-hand side of the autonomous system around each fixed point and constructing the corresponding Jacobian matrix. This Jacobian is an \( n \times n \) matrix, where \( n \) is the dimensionality of the phase space. The elements of the Jacobian matrix, for the $k$th fixed point, are given by:  
\begin{equation}
	J^{(k)}_{ij} = \left.\dfrac{\partial \dot{x}_i}{\partial x_j}\right|_{x_{c,k}}. \label{Jacobian}
\end{equation} 
After constructing the Jacobian matrix, the eigenvalues of the matrix can be computed. The stability of the fixed points is inferred from the sign of the real parts of these eigenvalues:
\begin{itemize}
	\item If all the real parts of the eigenvalues are negative, the corresponding fixed point is stable.
	\item If all the real parts are positive, the fixed point is unstable.
	\item If the real parts have mixed signs, the fixed point is a saddle point.  
	\item   If any real part vanishes, the standard linearization technique cannot determine the nature of the fixed point.  
\end{itemize}  
In cases where the real part vanishes, one can apply the center manifold theorem or numerically evolve the system around the critical points \cite{Boehmer:2011tp}. Note that in some cases, the stability of certain fixed points is indeterminate using the linearization method. For such cases, one has to rely on numerical evolution of the system to assess their stability. If the dynamical variables converge to the fixed point coordinates after numerically evolving the system around these critical points, the corresponding fixed point is deemed stable.

Though, the above mentioned scheme of finding the trajectories of the dynamical system and the fixed points works well for many systems, there are a few exceptions, such as
\begin{enumerate}
	\item when the phase space topology, as defined by Eq.~(\ref{pconstr}), is noncompact. In such cases, the dynamical variables can attain arbitrarily large values, resulting in an unbounded phase space. Consequently, fixed points located at infinity cannot be captured using the standard dynamical systems approach. 
	
	\item When some of the autonomous equations in the system become singular. This occurs when certain functions \( f_i(x_1, x_2, \cdots, x_n) \) (as defined in Eqs.~(\ref{gen-dyn-eqs})) diverge for finite, physically permissible values of the phase space variables. A common example arises from terms involving inverse powers, such as \( 1/x_i \), which render the equations ill-defined at \( x_i = 0 \).
\end{enumerate}
Remedies to tackle such pathological cases in dynamical analysis have been suggested in \cite{Bahamonde:2017ize, Bouhmadi-Lopez:2016dzw}. In brief, to tackle the noncompactness of the phase space, one can suitably map the phase space variables in terms of other dynamical variables which varies in a compact range adopting the most conventional one called Poincar\`e technique \cite{ELIAS2006305,BARREIRA20204416}. One then rewrites the autonomous equations in terms of the new dynamical variables of a compact phase space. We will see later that such situations arise in many inflationary dynamics, including the simplest case of slow-roll cold inflation. On the other hand, to deal with the divergences appearing in the autonomous equations, it is customary to suitably redefine $t^\prime$ or $N$ to get rid of the infinities. We will exploit such methods when we will deal with the dynamics of Warm Inflation in Sec.~\ref{warmi}.  

It is a common lore that inflationary trajectories in the phase space of its dynamical system often shows an attractor behaviour, which means that given a wide range of initial conditions of the dynamical variables the inflationary trajectories tend towards a specific region of the phase space (often tends toward and evolves around a fixed point). Our main aim would be to recognize such attractor solutions both in cold and warm inflation with slow-roll as well as constant-roll dynamics.

The material in this paper is organized as follows. In the next section \ref{crci}, we discuss the dynamics of CRCI for various values of the constant-roll parameter. Section \ref{warmi} presents the phase space analysis for constant-roll warm inflation (CRWI), wherein we compactify the phase space and regulate the dynamical system through a redefinition of time. In section \ref{wusr}, we extend the dynamical systems analysis to the case of warm ultraslow-roll inflation. Finally, we summarize our approach and highlight the key findings regarding inflationary dynamics in the concluding section \ref{sec:conclusion}.

\section{Dynamical analysis of cold inflation in the constant-roll regime}
\label{crci}

In standard cold inflation, the dynamics of the inflaton field \(\phi\) is governed by the Klein-Gordon equation  
\begin{equation}
	\ddot{\phi} + 3H\dot{\phi} + V_{,\phi} = 0, \label{eom_ci}
\end{equation}
where $V(\phi)$ is the inflaton potential and the subscript \(\phi\) indicates partial derivatives with respect to \(\phi\). Assuming the Universe is dominated by the inflaton field during cold inflation, the Friedmann equations take the form
\begin{eqnarray}
	3H^2 &= & \kappa^2 \left(\frac{1}{2}\dot{\phi}^2+V(\phi)\right)
	\label{first_friedmann_eq_CI},\\
	-2 \dot{H} &= &\kappa^2 \dot{\phi}^2
	\label{second_friedmann_eq_CI}\ .
\end{eqnarray}  
where \(\kappa^2 = 1/M_{\rm Pl}^2\), the reduced Planck mass is given by \(M_{\rm Pl}=1/\sqrt{8\pi G} \) where $G$ is the Newtonian gravitational constant. The Hubble slow-roll parameter slow-roll parameter $\epsilon_1$, defined as
\begin{equation}
	\epsilon_{1} = -\frac{ \dot{H}}{H^2}, 
	\label{eps1}
\end{equation}
is the primary parameter which specifies an inflationary phase (when $\epsilon_1\ll1$) as well as indicates the end of an inflationary phase (when $\epsilon_1\sim1$). Besides $\epsilon_1$, one can define other slow-roll parameters, though they do not play any significant role in the phase space analysis of the system. The energy density $\rho_\phi$ and the pressure \(P_{\phi}\) of the inflaton field,
\begin{eqnarray}
	\rho_{\phi} = \frac{1}{2}\dot{\phi}^2  + V(\phi),\quad\quad P_{\phi} = \frac{1}{2}\dot{\phi}^2  - V(\phi),
\end{eqnarray}
determine the equation of state \(\omega_{\phi}\) of the inflaton fluid as 
\begin{equation}
	\omega_{\phi} = \frac{P_{\phi}}{\rho_{\phi}} = \frac{\frac{1}{2}\dot{\phi}^2  - V(\phi)}{\frac{1}{2}\dot{\phi}^2  + V(\phi)} = - \frac{2 \dot{H}}{3 H^2} -1  = \frac{2}{3} \epsilon_{1} -1\ .
\end{equation}
This shows that during a slow-roll phase, when \(\epsilon_{1} \ll 1\), $\omega_\phi$ tends to \(-1\).

Constant-roll cold inflation (CRCI) was first introduced in \cite{Martin:2012pe, Motohashi_2015}, and the theory of CRCI was subsequently expanded and developed in \cite{Yi:2017mxs, Mohammadi:2022tmk, Shokri:2021zqw, Guerrero:2020lng}. Some of these papers have also carried out a dynamical system analysis of the inflationary process in constant-roll regime \cite{Lin:2019fcz,Mohammadi:2022tmk}. However, these previous studies were not solely dedicated to fully analyze the attractor nature of CRCI in a detailed manner. Here, we develop the general techniques and explore the dynamical system of CRCI. In CRCI all the essential dynamical equations of cold inflation are valid in addition with an extra constant-roll condition given by:  
\begin{equation}
	\ddot{\phi}=-3\beta H \dot{\phi}\label{cr_eq}\,,   
\end{equation}
where \(\beta\) is a parameter that generalizes both slow-roll and ultra-slow-roll inflation. This framework generalizes inflationary dynamics, as demonstrated in  Eq. (7) of Ref.~\cite{Motohashi_2015} where the parameter $\beta$ is related to $\alpha$ by the expression:
\begin{equation}
	\beta = \frac{1}{3} (\alpha + 3)\ .
\end{equation}
The condition in Eq.~(\ref{cr_eq}) modifies the cold inflation dynamics radically. For instance, as \(\beta \to 0\), the model approaches the slow-roll (SR) limit, whereas \(\beta \to 1\) corresponds to the ultra-slow-roll (USR) limit. This indicates that the larger the value of $\beta$, the more is the departure from the standard slow-roll dynamics. Moreover, unlike slow-roll cold inflation where one is free to choose the inflationary potential, the form of the potential during CRCI  gets fixed by the constant-roll dynamics. Depending on whether \(\beta\) is positive or negative, different forms of potentials can be derived that are suitable for CRCI dynamics. Above all, the graceful exit condition from a constant-roll dynamics can be different in cases with $\beta$ positive or negative as was first pointed out in Ref.~\cite{Biswas:2024oje}. 

Here, we will  represent dynamical analysis of CRCI using stream-plots as usage of autonomous equations turns out to be a difficult issue because of the constant-roll condition. The stream-plots will show the dynamical trajectories of the system and are capable of showing the attractor nature of the inflationary solution.\\
In the present section we will discuss:
\begin{enumerate}
	\item  the stability issue of CRCI dynamics for positive and negative values of $\beta$.
	
	\item The issue related to graceful exit in CRCI. 
	
	\item How does the attractor nature of CRCI differ from pure CI?
\end{enumerate}
In Ref.~\cite{Biswas:2024oje} the authors pointed out about the graceful exit problem in CRCI for both positive and negative values of $\beta$. In this article we verify the claims made in the reference. We show that in CRCI the attractor nature of inflation changes from cold inflation. 
\subsection{Case I: $\mathbf{\beta>0}$}

First we focus on the model for positive \(\beta\), i.e., \(\alpha > -3\). The corresponding potential for CRCI is given by \cite{Motohashi_2015}
\begin{equation}
	V(\phi) = V_{0}\left[\left(1-\beta\right)\cosh^{2}\left(3\kappa\sqrt{\frac{\beta}{6}}(\phi_{0}-\phi)\right) - 1\right]\label{cr_v_beta_positive}.
\end{equation}
Here $V_0\equiv 3M^2 M^2_{\rm Pl}$ is a positive constant, $M$ is a constant with the dimension of mass and \(\phi_0 \) is a constant value of the inflaton field at some fixed time.\footnote{The above form may appear different from the form of the potential appearing in Eq.~(2.26) of Ref.~\cite{Motohashi_2015}; however the present form can be derived using the identity \(\cosh(2\theta) = 2\cosh^2(\theta) - 1\) in Eq.~(2.26).}  
In Ref.~\cite{Motohashi_2015}, H. Motohashi et al. claimed that for a {potential} given in their Eq.~(2.26)-- which corresponds to Eq.~(\ref{cr_v_beta_positive}) in the present work-- the Hubble parameter in the context of CRCI is expressed as $H=M\tan[-3\beta M t]$. However, this expression is incomplete. The correct and complete form of the Hubble parameter is:
\begin{eqnarray}
	H=M\tan(-3\beta M t + C)\,,
	\label{hparam}
\end{eqnarray}
where \( C \) is an integration constant. Assuming \( M > 0 \), the Hubble parameter remains positive when the following condition is satisfied:
\begin{eqnarray}
	0 \le -3\beta M t + C < \frac{\pi}{2}\, .
	\label{tlim}
\end{eqnarray}
By restricting the argument of the tangent function to the first quadrant, the inequality becomes: 
\begin{eqnarray}
	\frac{1}{3\beta M}\left(C-\frac{\pi}{2}\right) < t \le \frac{C}{3\beta M}\,,
	\label{tlim1}
\end{eqnarray}
which can always be fulfilled if $C\sim \pi/2$ and adjusting the value of $M$. This condition specifies $H>0$. {Here, the constant $C$ is a physically meaningful parameter emerging from the constant-roll condition. Different choices of $C$ lead to different cosmological evolutions: for $C=0$, the Hubble parameter is negative near $t=0$, ruling out inflation, whereas for other values of $C$, one obtains $H>0$ near $t=0$, enabling an inflationary phase. In this sense, $C$ effectively encodes a freedom in redefining the origin of time.}
 As the scale-factor $a$ is always assumed to be greater than zero, implies $\dot{a}>0$. The authors of the above mentioned paper opines (after Eq.~(2.29) of their work) that ``Although this is a mathematically allowed solution, it has $\ddot{a}(t) < 0$. Therefore, it cannot describe an inflationary model in the usual sense.'' We show that such a conclusion is erroneous, there exists initial conditions and parameter values for which both  $\dot{a}>0, \ \ \ddot{a}>0$.\\
We have
\begin{equation}
	H\equiv \frac{\dot{a}}{a}=M \tan (-3\beta Mt+C)\,,
\end{equation}
which gives $\dot{a}=M \tan (-3 \beta Mt+C)a$ and consequently:
\begin{align}
	\ddot{a}=-aM^{2}[3\beta + (3\beta -1 )\tan^{2} (-3 \beta Mt+C)]\,.
\end{align}
In the limit $3\beta \to 0$, we see that the above expression can indeed produce $\ddot{a}>0$ during some inflationary epoch. Only when $3\beta > 1$ we do not find any possibility of inflation.

To analyze the dynamics associated with the potential given in Eq.~(\ref{cr_v_beta_positive}), we begin with the equation of motion for the inflaton field \(\phi\) as given in Eq.~\eqref{eom_ci}. Substituting the constant-roll condition from Eq.~\eqref{cr_eq}, the modified equation of motion for the inflaton field becomes:  
\begin{equation}
	3(1 -\beta) H \dot{\phi} + V_{,\phi} = 0\label{eom_cr}.
\end{equation}
It is important to note that the primary distinction between slow-roll and constant-roll inflation lies in the equation of motion for the inflaton field and the form of the potential. The remaining equations, specifically the first and second Friedmann equations, remain unchanged.

To understand the dynamics in this scenario, we define the dynamical variables as follows:  
\begin{equation}
	x = \pm \frac{\sqrt{V}}{\sqrt{3V_{0}}}, \quad y = \frac{\dot{\phi}}{\sqrt{6}\sqrt{V_{0}}}, \quad z = \frac{H}{\kappa \sqrt{V_{0}}}\label{crci_ol_var}.
\end{equation}
These dynamical variables are structurally similar to those defined for the earlier case and yield the same value of the slow-roll parameter used in the earlier case: $\epsilon_1=3y^2/(x^2 +y^2)$. Due to the different constants involved in the potential, we normalize the field and the Hubble parameter with the constant \(V_0\), thereby constructing dimensionless dynamical variables.  One must note that in the present case the potential in Eq.~(\ref{cr_v_beta_positive}) is not positive definite. This potential may admit negative values for some parameter choice and some values of the initial conditions. As because negative value of the scalar field potential can never produce inflation, we have worked with those initial conditions for which $V(\phi)$ is non-negative and consequently $x$ is properly defined everywhere in the phase space. From the form of the potential it is evident that when $\phi \sim \phi_0$, $V(\phi)$ turns negative and consequently in the present case we only analyze those dynamical regions where $x>0$ or $x<0$ as $x$ never crosses zero.

It is worth noting that, in this case, the constant-roll inflation condition reduces the second-order differential equation for \(\phi\) to a first-order equation. Hence, we can express the field \(\phi\) in terms of the dynamical variable using Eqs.~\eqref{cr_v_beta_positive} and \eqref{crci_ol_var}:  
\begin{equation}
	\kappa	\phi = \kappa\phi_{0} - \frac{1}{3} \sqrt{\frac{6}{\beta}} \ \cosh^{-1}\left(\sqrt{\dfrac{3 x^2 +1}{1-\beta}}\right)\ .
	\label{eq:phi}
\end{equation}
The above equation show that in our whole analysis $\beta<1$. The potential we are working with turns out to be purely negative for $\beta>1$ and consequently never produces inflation like solutions as in this regime $\ddot{a}<0$. Using the derivative of the potential:  
\begin{equation}
	V_{,\phi} = - 3 \kappa V_0 (1-\beta) \sqrt{\frac{\beta}{6}}  \sqrt{\left(\frac{2 (3 x^2 + 1)}{1-\beta} - 1\right)^2 -1}\,,
\end{equation}
the variable corresponding to $\dot{\phi}$ can be expressed using Eq.~\eqref{eom_cr}:
\begin{equation}
	y = \pm\frac{\sqrt{\beta}}{6 z}\sqrt{\left(\frac{2 (3 x^2 + 1)}{1-\beta} - 1\right)^2 -1}\ .
	\label{eq:y}
\end{equation}
The variable \(y\) is expressed in terms of \(z\) and \(x\) via the Hubble constraint relation: 
\begin{equation}  
	z^2 = x^2 + y^2\,,  
	\label{eq:const}
\end{equation} 
\begin{figure*}[t]
	\centering
	\includegraphics[scale=0.45]{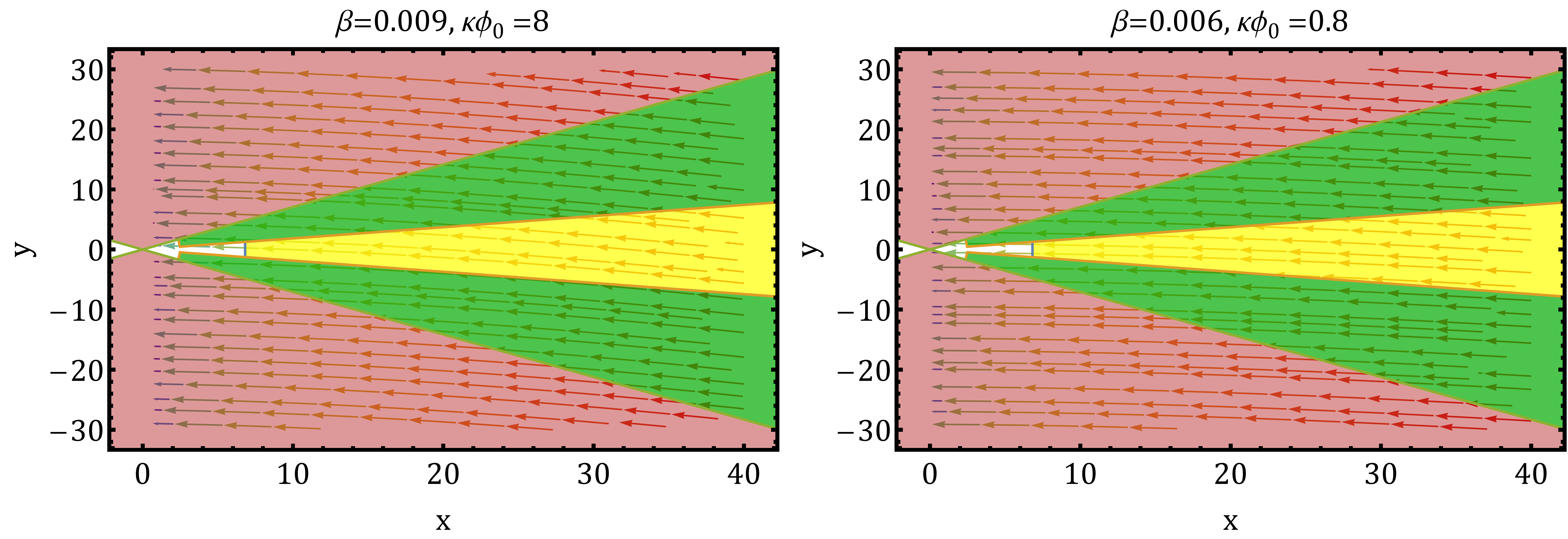}
	\caption{The streamline plots for the  field $(\kappa \phi, \dot{\phi}/\sqrt{V_0})$ in a specific $(x,y)$ range.
		The curves correspond to streamlines for different values of $\beta>0$ and field constants. The color code is  as follows: the green region represents regions where \(0<\epsilon_{1}<1\), in the pink region \(\epsilon_{1}>1\) and in the yellow region we have \(0< \epsilon_{1}<0.05\).}
	\label{fig:phase_constroll_cold}
\end{figure*}
which gives \(z\) in terms of \(x\) and $y$.  We have produced the streamline plots for the field $(\dot{\phi}/\sqrt{V_0}, \kappa \phi)$ in some specific range of $x$ and $y$. The stream lines are the flow lines for the vector field $(\kappa \phi, \dot{\phi}/\sqrt{V_0})$. The stream-plots are shown in Fig.~[\ref{fig:phase_constroll_cold}] for different values of \(\beta\) and $\kappa \phi_0$.  Each stream-plot is divided into distinct colored regions: the green region represents \(0 < \epsilon_1 < 1\), the pink region represents \(\epsilon_1 > 1\), and the yellow region represents \(0 < \epsilon_1 < 0.05\).  On the boundary curve, differentiating the green and the pink regions, $\epsilon_1=1$.

The flow-line plots are done for $x>0$, and the flow lines are towards the minimum of the potential (towards $x=0$).
When $\beta$ is reasonably small (near the slow-roll limit), for both the two plots in Fig.~[\ref{fig:phase_constroll_cold}], it is seen that the streamlines originating in the yellow region, with negative as well as positive values of \(\dot{\phi}\) in the right side of the plots, have initial value of $\epsilon_1$ to be very small. Most of these flow lines leave the yellow region and move towards the pink region through the green patch. These flows represent potential inflationary flows. As soon these lines come near the pink region, inflation ends. It is seen that near the slow-roll limit, CRCI still remains an attractor solution. 
In the present case we do not employ the autonomous equations to figure out the inflationary dynamics as there is only one independent variable and the phase space is one dimensional. The streamline plots in the present case do not give us any information about the fixed points in CRCI. 

\begin{figure}[t]
	\centering
	\includegraphics[scale=0.4]{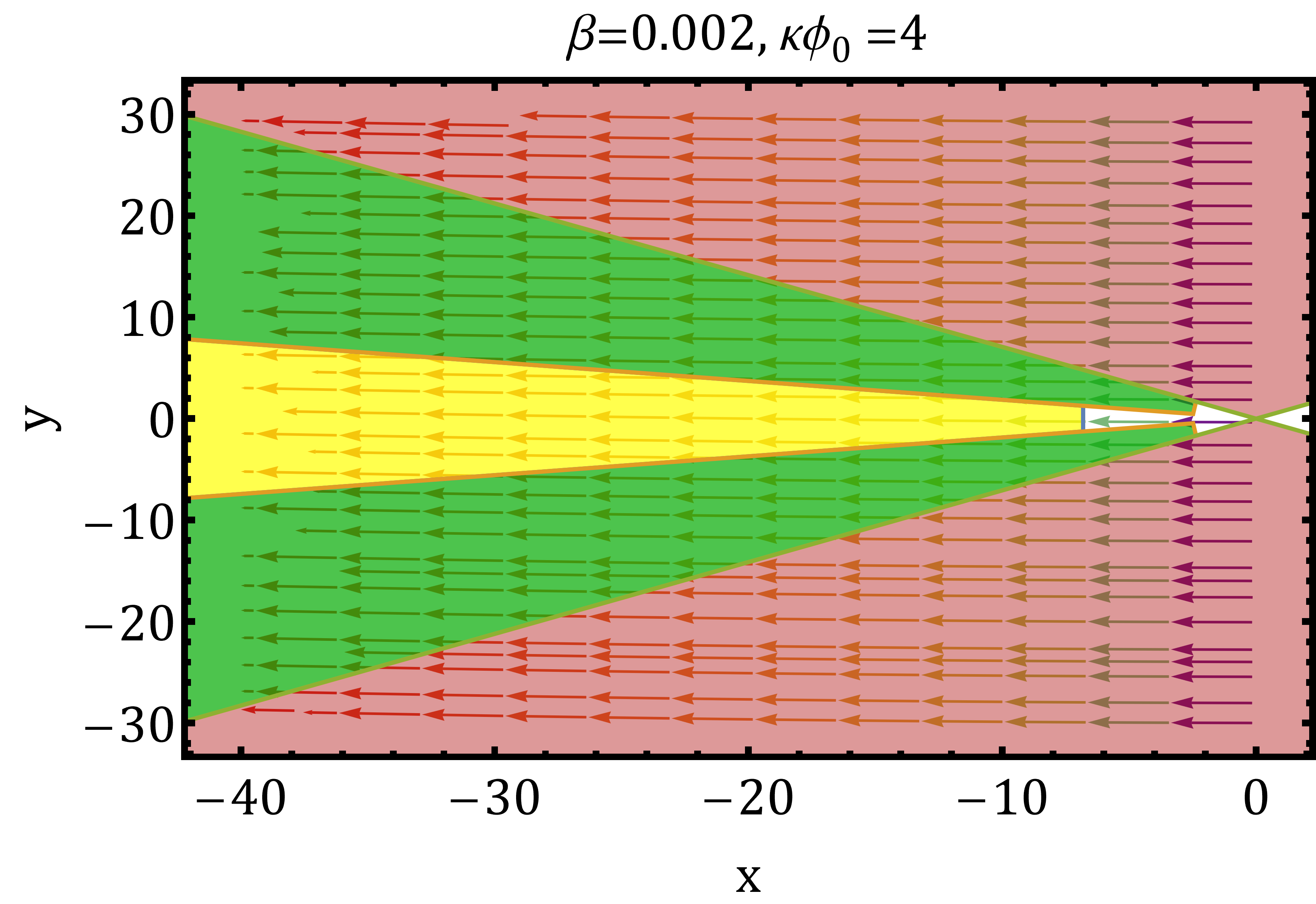}
	\caption{The streamline plots for the  field $(\kappa \phi, \dot{\phi}/\sqrt{V_0})$ in a specific $(x,y)$ range.
		Streamlines for $\beta = 0.002$, illustrating that they remain confined within the inflation region and do not exit. The color coding is the same as the previous figures.}
	\label{fig:phase_constroll_cold_no_end}
\end{figure}

For a complete description of CRCI dynamics one should note that in this case one encounters eternally inflating trajectories in the stream-plots for appropriate parameters and range of the variables. In Fig.~[\ref{fig:phase_constroll_cold_no_end}], we plot the stream lines for negative values of $x$. In this case one is moving away from $V(\phi)=0$. The plot shows that there are stream lines which will eternally be inside the yellow region, if initially they start from the yellow region. For these kind of dynamics the stream lines never get a chance to reach the pink region.  This is a generic feature of most CRCI models.

In the present case the standard autonomous system of equations are not required as the system is solvable algebraically
and the flow dynamics specified by $(\kappa \phi, \dot{\phi}/\sqrt{V_0})$, as shown in Fig.~[\ref{fig:phase_constroll_cold}], is sufficient in the present case and shows the qualitative behavior of the system exhaustively. 
\subsection{Case II: $\mathbf{\beta<0}$}

In this section, we will be investigating the scenario where the constant roll parameter \(\beta\) can take negative values. Hence, the corresponding constraint equation becomes:  
\begin{equation}
	\ddot{\phi}=3\tilde{\beta}H\dot{\phi}\label{crci_nb}\ ,
\end{equation}
where we redefined \(\beta\) as \(\tilde{\beta} = -\beta\). It has been shown in Ref.~\cite{Motohashi_2015} that negative \(\beta\) modifies the form of the potential. Therefore, the corresponding potential as mentioned in Ref.~\cite{Motohashi_2015} is\footnote{The above form may appear different from Ref.~\cite{Motohashi_2015}; however, the present form has been derived using the identity \(\cos(2\theta) = 1-2\sin^2(\theta) \) of Eq. (2.22).}
\begin{equation}  
	V(\phi) = V_{0}\left[1 - (1+\tilde{\beta})\sin^2\left(3\kappa\sqrt{\frac{\tilde{\beta}}{6}}(\phi+\phi_{0})\right)\right]\label{V_crci_nb} \ . 
\end{equation}  
As we introduce the redefined constant roll parameter \(\tilde{\beta}\), the corresponding equation of motion of the inflaton field, assuming the constant-roll constraint, becomes:  
\begin{equation}  
	3H\dot{\phi}(1+\tilde{\beta}) + V_{,\phi} = 0\label{crci_eom} \ .  
\end{equation}  
To proceed with the dynamical analysis of the system, as carried out in the previous case, we use similar dimensionless variables as defined in Eq.~(\ref{crci_ol_var}), as both cases have a similar structure.  In the present case also the potential $V(\phi)$ admits negative values and consequently we have to work either with positive values or negative values of $x$. As inflation always occurs when $V(\phi)>0$ we do not work with those initial conditions which produce negative scalar field potential.

Due to the definition of the variables, the field \(\phi\) can be expressed as:  
\begin{equation}
	\kappa\phi=\frac{1}{3}\sqrt{\frac{6 }{\tilde{\beta}}}\sin^{-1}{\left(\sqrt{\frac{1-3x^2}{1+\tilde{\beta}}}\right)}-\kappa\phi_0\label{crci_nb_phi}\ .
\end{equation}
Similarly, the gradient of the potential becomes:
\begin{equation}
	V_{,\phi}=\pm 3\kappa V_{0}(1+\tilde{\beta})\sqrt{\frac{\tilde{\beta}}{6}}\sqrt{1-\left(\dfrac{2(1-3x^2)}{(1+\tilde{\beta})}-1\right)^{2}}\label{crci_nb_dV}, 
\end{equation}
and the time derivative of the field \((\dot{\phi} \propto y)\) can be expressed as:
\begin{equation}
	y=\pm\frac{\sqrt{\beta}}{6z}\sqrt{1-\left(\dfrac{2(1-3x^2)}{(1+\tilde{\beta})}-1\right)^{2}}\label{crci_nb_y}\ .
\end{equation}
It is important to note that from Eq.~(\ref{crci_nb_phi}), the argument of the inverse sin function must lie within the range \([-1, 1]\). This imposes the constraint on \(x\):
$$-\frac{1}{\sqrt{3}}<x<\frac{1}{\sqrt{3}}\,.$$
The above condition shows one of the differences of CRCI, with negative $\beta$, with respect to the case of positive $\beta$. In the present case the range of $x$ is bounded. This also puts a bound on the maximum value of $y$, for $\tilde{\beta}=0.0067$. The bounded interval for $y$ is $(-0.4,0.4)$.  

\begin{figure*}[t]
	\centering
	\includegraphics[scale=0.45]{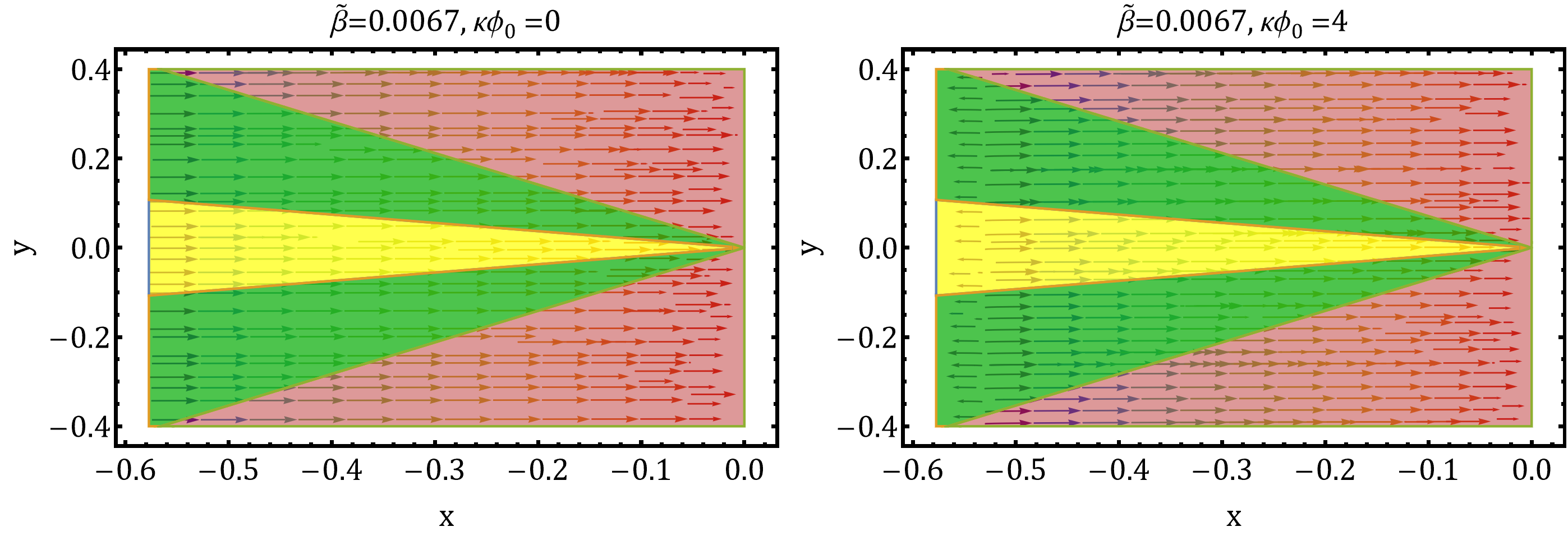}
	\caption{The streamline plots for the field $(\kappa \phi, \dot{\phi}/\sqrt{V_0})$ in a specific $(x,y)$ range. The curves correspond to streamlines for $\beta<0$. For these plots $\tilde{\beta}=0.0067$ and the field constant vary.  Here green region corresponds to regions where \(0.1<\epsilon_{1}<1\), pink region corresponds \(\epsilon_{1}>1\), and yellow region corresponds to \(0< \epsilon_{1}<0.1\).}
	\label{fig:phase_constroll_cold_nb}
\end{figure*}

As in the case of positive $\beta$, we have constructed the stream-plots showing the nature of streamlines corresponding to the components $(\kappa \phi, \dot{\phi}/\sqrt{V_0})$ by varying the independent variable $x$. The streamline plots are shown in Fig.~[\ref{fig:phase_constroll_cold_nb}]. For this analysis, we have chosen $\tilde{\beta} = 0.0067$, as this value corresponds to a spectral index $n_s = 0.96$, consistent with the value obtained in \cite{Motohashi_2015}.

In the case of negative \(\beta\), the stream-plots exhibit a similar attractor like solution as in the previous case, however this time we have only confined the dynamics only in the negative $x$ region. The stream lines plotted in  Fig.~[\ref{fig:phase_constroll_cold_nb}] compel the system to move towards potential zero. It is seen in this case that all the streamlines originating on the left most side must be streaming in such a way that for all of them $\dot{\epsilon_1}>0$.
This behavior aligns with the findings of Ref.~\cite{Biswas:2024oje}. 

Like the previous case for $\beta>0$, in this case also we get the phase space behavior algebraically. It is important to note that the extra constant-roll condition permits the existence of inflationary phase  in both the cases discussed in this section, for positive and negative values of $\beta$. In both the cases it is observed that the attractor nature of inflation has been affected by the constant-roll condition. When we compare our results with the result of cold inflation (as given in the previous section) we see that only a part of the phase space trajectories end up in inflationary phase in the present case whereas in the case of cold inflation mostly all trajectories were attracted towards some inflationary phase. 

Before we finish our discussion on CRCI it is important to specify that the authors in Ref.~\cite{Lin:2019fcz, Morse:2018kda} have studied the large $\beta$ behavior of CRCI models and have shown that these models do not in general show attractor behavior. They conclude that only small $\beta$ solutions, which specify slow-roll limit of CRCI, can only produce attractor like solutions. In the present paper we have not delved into the large $\beta$ limit as our primary aim was to unravel the attractor nature of CRCI. 

{The results presented in this section can be compared with the result of another contemporary work in this field. In Section~2.5 of Ref.~\cite{Motohashi:2025qgd}, the authors state that for the range \(-{3}/{2} < \beta < 0\), the background solution is an attractor. Since we have worked with \(\beta = -0.0067\), which lies within this interval, our results are fully consistent with their analysis: we also obtain an attractor solution. Moreover , in Section~2.4 of the above reference, one of the values of \(\beta\) considered by the authors is \(0.01\), which is in close proximity to one of the values we have used in the previous subsection, namely \(\beta = 0.009\). These facts show that our parameter choices are consistent with the latest work on this field.}

\section{Constant-roll warm inflation (CRWI)}
\label{warmi}

The warm inflation scenario arises when the inflaton field dissipates energy into lighter degrees of freedom at a rate exceeding the Hubble expansion rate. This dissipation mechanism ensures that the generated particles thermalize efficiently, creating a radiation bath. During this epoch, these light particles can be modeled as radiation, with the radiation energy density given by:  
\begin{equation}
	\rho_r = \frac{\pi^2 g_* T^4}{30},
\end{equation}
where \(\rho_r\) is the radiation energy density, \(g_*\) is the effective number of relativistic degrees of freedom, and \(T\) is the temperature of the generated particles.  

The evolution of the inflaton field \(\phi\) and the radiation energy density \(\rho_r\) in a spatially flat FLRW background is described by the following equations:  
\begin{eqnarray}
	\ddot{\phi}+3H\dot{\phi}+  V_{,\phi}&=& -\Upsilon(\phi,T) \dot{\phi}\ , \label{wi_field_eom}\\
	\dot{\rho}_{r}+4H\rho_{r}&= &\Upsilon(\phi,T) \dot{\phi}^2\ . \label{wi_rad_eom}
\end{eqnarray}
Here $\Upsilon$ is the dissipation rate, the rate at which the inflaton sector dissipates energy to the radiation bath. One can get a form of $\Upsilon$ from the microphysics of the system. The form of $\Upsilon$ depends upon the nature of particles produced via the decay of the inflaton. If these decays (to multiple types of particles) happen near a thermal equilibrium one may expect $\Upsilon$ to be a function of the ambient temperature of the radiation bath and the inflaton field strength. As $\Upsilon$ is related to the decay rate of the inflaton, we must obviously have the following dynamical constraint:
\begin{equation}
	\Upsilon \ge 0\,.
\end{equation}
When $\Upsilon=0$, the radiation and the scalar field sector decouples. 
In general the dissipative factor can be expressed as \cite{Kamali:2023lzq,Bastero-Gil:2016qru,Berera:2008ar,Berghaus:2019whh}: 
\begin{equation}
	\Upsilon = C_{1}T^c\phi^pM^{1-p-c}\label{gen_upsilon} \ ,
\end{equation}
where $C_{\Upsilon}$ is a dimensionless constant, \(c, p\) are the free parameters of the model and \(M\) is the mass dimensional constant. In the literature the following forms of the $\Upsilon$ have been studied extensively \cite{Kamali:2023lzq,Bastero-Gil:2016qru,Berera:2008ar,Berghaus:2019whh}. 
The energy density and the pressure of the inflaton field are: 
\begin{equation}
	\rho_{\phi} = \frac{1}{2}\dot{\phi}^2+V(\phi), \quad P_{\phi} = \frac{1}{2}\dot{\phi}^2  - V(\phi)\ .
\end{equation}
The first and second Friedmann equations are: 
\begin{eqnarray}
	3H ^2  &=&  \kappa^2 (\rho_{r} + \rho_{\phi})\ , \label{first_fried_wi} \\
	2 \dot{H} &=&   -\kappa^2 \left(\dot{\phi} ^2 + \frac{4}{3} \rho_{r} \right)\ .
\end{eqnarray}
Given a form of $\Upsilon$ one can see whether one can get a state of inflation, during which the slow-roll conditions are maintained. During such a slow-roll phase the radiation bath maintains a dynamic equilibrium so that one can attach a temperature $T$ of the radiation bath. Warm inflation happens as long as $T > H$. To ensure that the warm inflation field can initiate the consistent inflationary epochs, a number of slow roll parameters are usually defined as: 
\begin{equation}
	\epsilon_{1} =  - \frac{\dot{H}}{H^2}, \quad \eta_{1}   = \dfrac{1}{\kappa^2 (1+Q)}\dfrac{V_{,\phi \phi}}{V} , \quad \beta = \dfrac{\Upsilon_{,\phi} V_{,\phi}}{\kappa^2 V \Upsilon }\,. 
\end{equation}
{Additionally, in this case we do not assume that the radiation bath has a fixed temperature initially}. However, it may happen the system ultimately settles down to a phase where $\rho_r$ becomes a constant, with $T>H$, and we may interpret the dynamics of that phase to be the dynamics of a warm inflationary phase, if the slow-roll conditions are satisfied. If cosmological dynamics does not produce a phase with nearly a constant radiation energy density then we do not get a proper warm inflationary phase. Moreover, throughout the dynamical evolution of the system, the radiation energy density must be positive i.e. $\rho_{r}>0$ and the system must be thermally stable: 
\begin{equation}
	\frac{|\dot{\rho}_r|}{H\rho_r} \ll 1\ , \quad\text{or equivalently}\quad \frac{|\dot{T}|}{HT}\ll 1.  \label{thermal_stab_constraint}
\end{equation}


\label{crwis}

In this section, we explore the constant-roll scenario in the framework of warm inflation. The constraint equation governing the constant-roll warm inflation, remains the same as introduced in Eq.~\eqref{cr_eq}:  
\begin{equation}
	\ddot{\phi}=-3\beta H \dot{\phi} \, .
	\nonumber    
\end{equation}
In CRWI the Friedmann equations in Eq.~\eqref{first_fried_wi} remain unchanged, however, the scalar field equation becomes a first order differential equation:   
\begin{eqnarray}
	3H\dot{\phi}\left[(1+Q)-\beta\right]=-V_{,\phi}\,. 
	\label{const_wi_field}
\end{eqnarray}
Here $Q=\Upsilon/3H$ as defined in the previous section on warm inflation. The constant-roll condition always acts like a dynamical constraint and turns the second order differential equation for the inflaton into a first order differential equation. This fact has interesting consequences as was observed in the case of CRCI where the dynamics of the system could be obtained from purely algebraic equations. We will like to see how warm inflation dynamics is affected by the constant-roll constraint. As in CRCI, in this case also we have different kind of inflaton potentials for positive and negative values of $\beta$.     

It is known that in CRWI if $Q$ is a function of the inflaton field and the temperature then the dynamical system becomes ill 
defined \cite{Biswas:2024oje} if one demands CRWI to be taking place near a dynamic thermal equilibrium. This difficulty arises because in such a case the thermal stability condition and the constant-roll condition combines to produce a scalar field solution which may not respect the Friedmann equations.  Consequently, CRWI is a highly constrained system where only constant $Q$ is allowed. One may also work with a temperature dependent $Q$, but as thermal stability is assumed, we will have temperature to be approximately constant throughout the inflationary phase and consequently temperature dependent $Q$ also behaves a constant $Q$ system. These points are elaborately explained in Ref.~\cite{Biswas:2024oje}. Henceforth we address the various cases of CRWI corresponding to positive and negative values of $\beta$ respectively. We will only analyze the inflationary dynamics assuming constant $Q$. 
\subsection{CRWI with $\beta>0$ and constant $Q$}

\begin{figure*}[t!]
	\centering
	\includegraphics[scale=0.5]{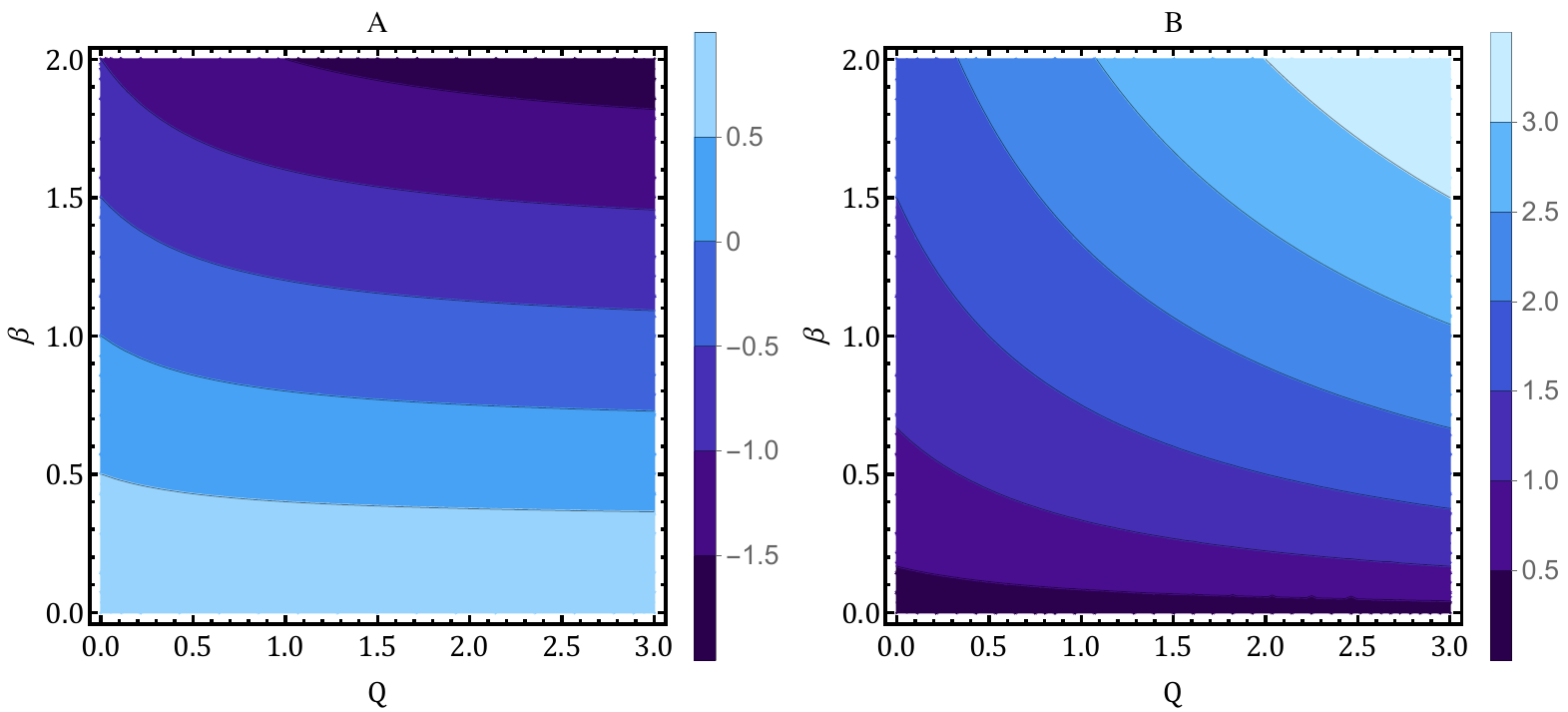}
	\caption{The variation of \(\mathcal{A}\) and $\mathcal{B}$ in the parameter space of \((Q, \beta)\) when $\beta>0$. }
	\label{fig:a_plot}
\end{figure*}

Constant-roll conditions affect the graceful exit problem in most inflationary theories. It is generally seen that many inflaton potentials which can cause inflation like behavior are not suitable as with those potentials the inflationary phase never ends.
Graceful exit within this framework, for \(\beta > 0\), can be achieved with the inflaton potential given by \cite{Biswas:2024oje}:  
\begin{equation}
	V(\phi)=V_{0}\left[\mathcal{A} \cosh^2{\left(\mathcal{B}\kappa(\phi_0-\phi)\right)}-1\right]\label{CRWI_V},
\end{equation}
where \(\mathcal{A} \equiv \dfrac{2(1+Q) - 2\beta - 3Q\beta}{2(1+Q)}\), \(\mathcal{B} \equiv 3\sqrt{\frac{(1+Q)\beta}{6}}\), and \((\phi_0, V_0)\) are constants. It is observed that the constants $\mathcal{A}, \mathcal{B}$ depends on the values of $Q$ and $\beta$. The variation of \(\mathcal{A}\) and \(\mathcal{B}\) is depicted in Fig.~[\ref{fig:a_plot}] for various values of $Q,\,\,\beta$.   
From the plots, it is evident that \(\mathcal{A}\) can take negative values for \(\beta > 1\) and \(Q > 0\), while \(\mathcal{B}\) remains positive. However, \(\mathcal{B}\) becomes complex for negative \(\beta\), so negative values of \(\beta\) are excluded. 

To explore the dynamical scenario in this case, we define the dimensionless dynamical variables in a manner we have done so far for other cases:   
\begin{equation}
	x^2 = \frac{{V}}{{3V_{0}}},\quad y=  \frac{\dot{\phi}}{\sqrt{6}\sqrt{V_{0}} },  \quad z = \frac{H}{\kappa \sqrt{V_{0}}} \label{var_Crwi}.
\end{equation}
The other variables can be expressed in terms of the primary variables as:  
\begin{equation}
	\Omega_{r} = \frac{\kappa^2 \rho_{r}}{3 H^2} = \frac{ \rho_{r}}{3 z^2 V_0}, \quad \Omega_{\phi} = \frac{\kappa^2 \rho_{\phi}}{3H^2} = \frac{x^2 +y^2}{z^2} \ .
\end{equation}
With these variables, the first Friedmann equation becomes:   
\begin{equation}
	1 = \Omega_{\phi}  + \Omega_{r} . \label{cwi_constraint}
\end{equation}
Similarly, the other important quantities, such as the slow-roll parameter and the thermal stability parameter,  become:  
\begin{equation}
	\epsilon_{1} = \frac{-y^2 - 2 z^2 + 2 x^2}{z^2}, \quad \frac{\dot{\rho}_{r}}{4\rho_{r} H} = \frac{3 Q y^2}{2(z^2-x^2-y^2)}-1 \ . 
\end{equation}
One can see that the inflaton potential used in the present context is not positive definite but for inflation we will be working with a class of initial conditions and a particular set of parameter values for which $V(\phi)>0$. As because $V$ can be negative, we will have to work only with the positive or negative branch of $x$. Here we work with positive values of $x$. The primary variables \((x, z)\) can take any values within the range \([0, \infty)\). To keep track of the behavior of the variables at infinity we compactify the space spanned by these variables and define:   
\begin{equation}
	\bar{x} = \dfrac{x}{\sqrt{1+x^2}},  \quad  \bar{z}\equiv\dfrac{z}{\sqrt{1+z^2}} \label{compactification1}\,.
\end{equation}
These transformations map the infinite interval to a finite interval \([0, 1)\). 

Using the variables defined in this section we can now construct the autonomous system of equations. From our previous discussion we know that in the present case we have  two independent variables which can describe the dynamics of the system. The autonomous system of equations corresponding to the relevant variables \((\bar{x}, \bar{z})\) are:   
\begin{eqnarray}
	\bar{x}'&=& -\dfrac{(1-\bar{x}^2)^2\sqrt{\frac{1}{\mathcal{A}}\dfrac{2\bar{x}^2+1}{1-\bar{x}^2}}\sqrt{\frac{1}{\mathcal{A}}\dfrac{2\bar{x}^2+1}{1-\bar{x}^2}-1}}{\bar{x}} \times \nonumber\\
    && \left({y}\mathcal{A}\sqrt{(1+Q)\beta}\right) \ ,\label{auto_eq_xb}\\
	\bar{z}'&=& \left(1-\bar{z}^2\right)^{\frac{3}{2}}\left(-y^2-2\dfrac{\bar{z}^2}{1-\bar{z}^2}+2\dfrac{\bar{x}^2}{1-\bar{x}^2}\right) \ \label{auto_eq_zb}.
\end{eqnarray}
Here, the derivative is defined as \(()' \equiv \frac{d()}{\kappa \sqrt{V_0} dt}\), and \(y\) can be expressed in terms of \((\bar{x}, \bar{z})\) using the relation 
\begin{equation}
	y=\mp\frac{2}{\sqrt{6}}\frac{\sqrt{1-\bar{z}^2}}{\bar{z}}\frac{\mathcal{A}  \sqrt{\frac{(1+Q)\beta}{6}}}{\left((1+Q)-\beta\right)}\sqrt{\frac{1}{\mathcal{A}}\dfrac{2\bar{x}^2+1}{1-\bar{x}^2}}\sqrt{\frac{1}{\mathcal{A}}\dfrac{2\bar{x}^2+1}{1-\bar{x}^2}-1} \label{crwi_y} \,.
\end{equation}
Since the autonomous system of equations requires only two variables to fully describe the dynamics, the resulting phase space is two-dimensional.
One must note that the first autonomous equation above explicitly  depends on $y$ which has two branches, one positive and the other negative, clearly seen from 
Eq.~\eqref{crwi_y}. In the present case we have worked with the positive branch of $y$ as the results can be interpreted with ease in such a case.
\begin{table}[t!]
	\centering
	\begin{tabular}{|c|c|c|c|c|c|}\hline
		
		Points&  \((\bar{x}_*, \bar{z}_*) \)&  \(\Omega_\phi \)&  \(\epsilon_1 \)& \(\frac{\dot{\rho}_{r}}{H \rho_{r}}\) & Stability \\ \hline 
		
		$P_{1}$ & $(0,0)$ & $\infty$ & $\infty$ & $-\frac{3 Q}{2}-1$ & Unstable\\
		
		\hline
		
		$P_{2}$ & $( 1, 1)$ & Indeterminate & $2$ & $-1$ & Unstable\\

		\hline
		
		%
		
		$P_{3}$ & $(\rm{Any}, 1)$ & $0$ & $2$ & $-1$ & Unstable\\
		
		\hline
		
	\end{tabular}
	\caption{The critical points and their nature for CRWI Model when $\beta>0$. }
	\label{tab:warm_constant}
\end{table}

It is important to note that the differential equation system diverges for \((\bar{x} = 0, 1)\) and \(\bar{z} = 0\). To address these divergences, we redefine the time variable as follows:   
\begin{equation}
	\kappa \sqrt{V_0} dt  \to \kappa \sqrt{V_0} \bar{z}^2 \bar{x} (1-\bar{x}^2)^2 d{t} \ .
\end{equation}
The modified autonomous equations will become: 
\begin{eqnarray}
	\bar{x}'&= &-\bigg({y}\bar{z}^2\mathcal{A}\sqrt{(1+Q)\beta}(1-\bar{x}^2)^4\sqrt{\frac{1}{\mathcal{A}}\dfrac{2\bar{x}^2+1}{1-\bar{x}^2}}\sqrt{\frac{1}{\mathcal{A}}}\nonumber \\
    && \dfrac{2\bar{x}^2+1}{1-\bar{x}^2} \bigg)-1  \ ,\\
	\bar{z}'&=& \bar{z}^2 \bar{x} \left(1-\bar{z}^2\right)^{\frac{3}{2}}\left(-y^2-2\dfrac{\bar{z}^2}{1-\bar{z}^2}+2\dfrac{\bar{x}^2}{1-\bar{x}^2}\right)\\ \nonumber && \times (1-\bar{x}^2)^2 \ .
\end{eqnarray}
Here, the derivative is defined as \(()' \equiv \frac{d()}{\kappa \sqrt{V_0} d{t}}\).
To determine the qualitative behavior, we compute the critical points corresponding to the redefined equations. These critical points are summarized in Tab.~[\ref{tab:warm_constant}]. The system yields three critical points, none of which depend on any model parameters.

At point \(P_1\), both coordinates vanish, resulting in an indeterminate fractional energy density. Additionally, the slow-roll parameter becomes very large, rendering this point unstable.

At the point \(P_2\) the field fractional energy density becomes indeterminate, and the slow-roll parameter exceeds 1. Upon evaluating their stability, the point \(P_2\) exhibits unstable behavior for any choice of model parameters.  Here the point \(P_3\) is not representing any specific point, as \(\bar{x}\) can take any value. Here $P_3$ actually represents a line of critical points where \(\bar{x} \ne 1\). This line of critical points is generally unstable.  

\begin{figure}[t]
	\centering
	\includegraphics[scale=0.25]{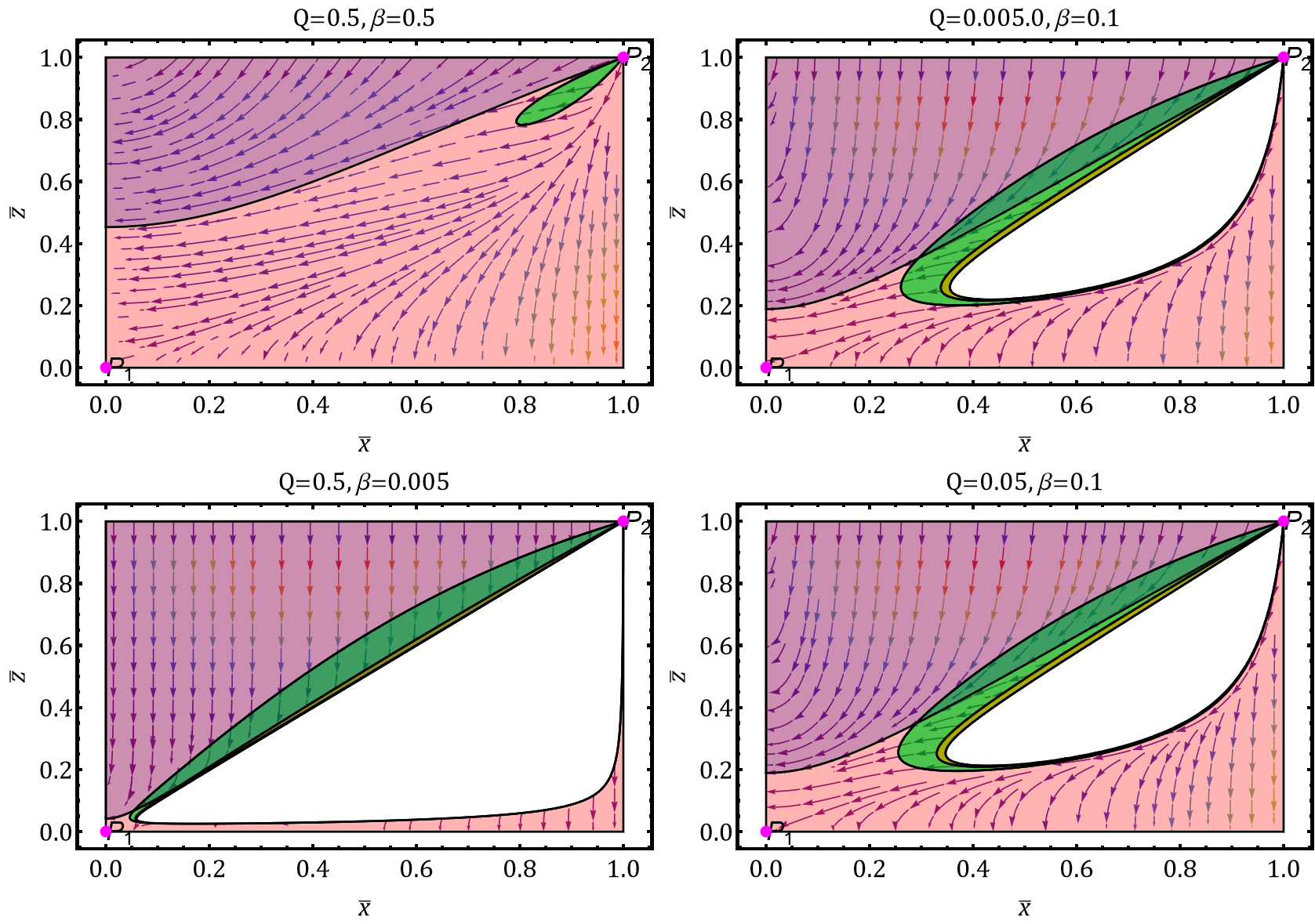}
	\caption{The phase space is corresponding to CRWI when $\beta>0$. The color scheme highlights the different dynamical regions: pink (\(\epsilon_1 \geq 1\)), yellow region  (\(0 \leq \epsilon_1 \leq 0.1\)), green region  (\(0.1 < \epsilon_1 < 1\)), blue (thermally stable region), and violet (overlap of red and blue regions).}
	\label{phase_crwi}
\end{figure}

The phase space is depicted in Fig.~[\ref{phase_crwi}] for different values of \((Q, \beta)\) in the parameter space of \((\bar{x}, \bar{z})\). The phase space is illustrated using distinct color regions: pink represents $\epsilon_1 \geq 1$, green corresponds to $0.1 < \epsilon_1 < 1$, yellow indicates $0 \leq \epsilon_1 \leq 0.1$, and violet denotes the thermally stable region. The white region represents an unphysical domain where $\epsilon_1$ becomes negative.

For $(Q = 0.5, \beta = 0.5)$, no physically viable dynamics exist. Conversely, for $\beta \ll 0.1$, as shown in the bottom left panel, trajectories originating from the line of critical points ($\bar{z} =1, \bar{x} \ne 1$) maintain thermal stability and are drawn towards the green region, eventually entering the yellow region, where the slow-roll condition $0 \leq \epsilon_1 \leq 0.1$ is satisfied. These physically viable trajectories exit the inflationary region and transition into the pink region maintaining thermal equilibrium, marking the end of inflation. Something similar happens in the lower right figure, but in this case the trajectories from the yellow region does not always remain in thermal equilibrium as they flow towards the pink region. In an overall manner, the dynamics of these trajectories confirm  that CRWI  can occur although the phase space of these processes is much more constrained and complicated.


\begin{figure}[t!]
	\centering
	\includegraphics[scale=0.4]{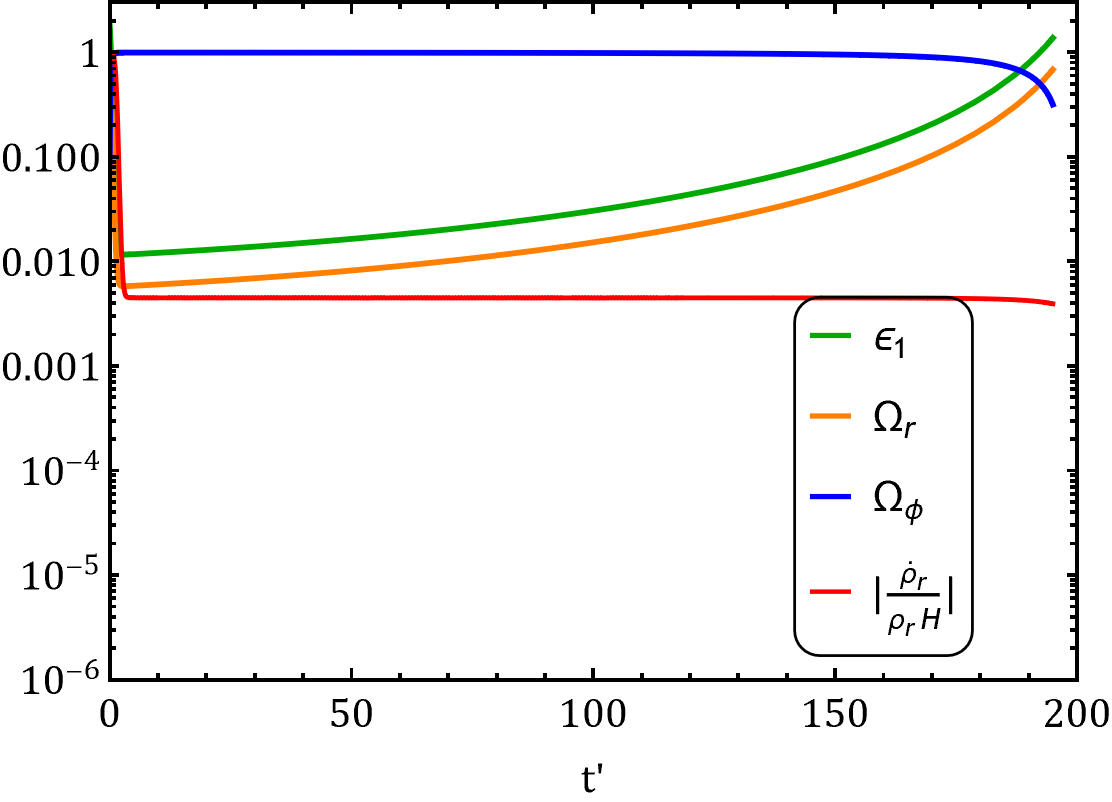} 
	\includegraphics[scale=0.65]{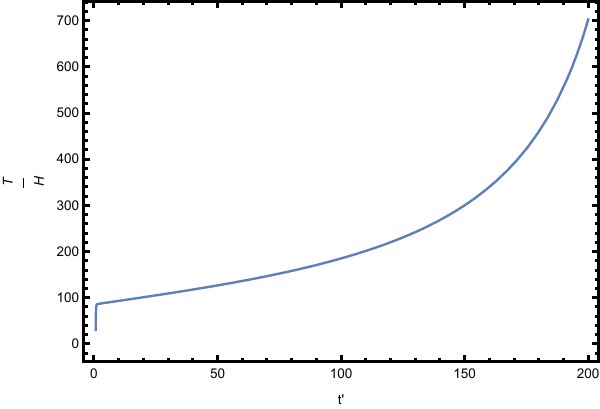}
	\caption{The evolution of CRWI corresponding  variables and plot of $\frac{T}{H}$ of CRWI corresponding to $(Q=300, {\beta} = 3\times10^{-3})$ and $V_{0}=10^{-6} 
		M^{4}_{\rm Pl}$.}
	\label{evo_crwi}
\end{figure}

If we follow one of the trajectories in the phase space we can construct the explicit inflationary dynamics which correspond to that trajectory. Here we plot 
the evolution of the cosmological parameters against the redefined time variable \({t'}\) for $Q = 300,\, \beta = 3\times10^{-3}$ in left panel of Fig.~[\ref{evo_crwi}]. From the plot we can see that during the initial phase, the field energy density dominates over the radiation energy density, and the corresponding slow-roll parameter remains less than one. At around $t^\prime \sim 200$ the radiation energy density starts to dominate and inflation ends. At around this time we observe $\epsilon_1 \sim 1$. On the right hand panel of  Fig.~[\ref{evo_crwi}], we show the evolution of $T/H$ during the inflationary period. It is clearly seen that during the inflationary period we have $T>H$ showing that thermal fluctuations dominate over quantum fluctuations.  

As the radiation energy density starts to dominate, the slow-roll parameter exceeds 1, signaling the system's exit from the inflationary phase after \({t'}=195\) which corresponds to about 87 $e$-folds. It must be noted that by tuning the initial conditions one can reduce the number of $e$-folds. This tuning requires a detailed search in the initial condition space and parameter space of the theory. In this paper we present the method in principle and show that a through numerical stability analysis is possible in highly constrained inflationary systems. To calculate the number of e-folds, we use the formula: $N = \int H \, dt,$
which, in our case, can be expressed as:
\begin{align}
	N = \int H \, dt  = \int \kappa \sqrt{V_{0}} z \, dt 
	= \int \frac{\bar{z}}{\sqrt{1 - \bar{z}^{2}}} \, dt'. \nonumber
\end{align}
Here, \(\bar{z}\) is obtained by solving Eqs. (\ref{auto_eq_xb}--\ref{auto_eq_zb}) as a function of \(t'\). By integrating over the entire duration of inflation in terms of \(t'\), we determine the total number of $e$-folds. Throughout the evolution, the modulus of the thermal stability parameter remains less than 1, ensuring thermal stability during the entire process.

\subsection{CRWI with $\beta<0$ and constant $Q$}

\begin{figure*}[t!]
	\centering
	\includegraphics[scale=0.5]{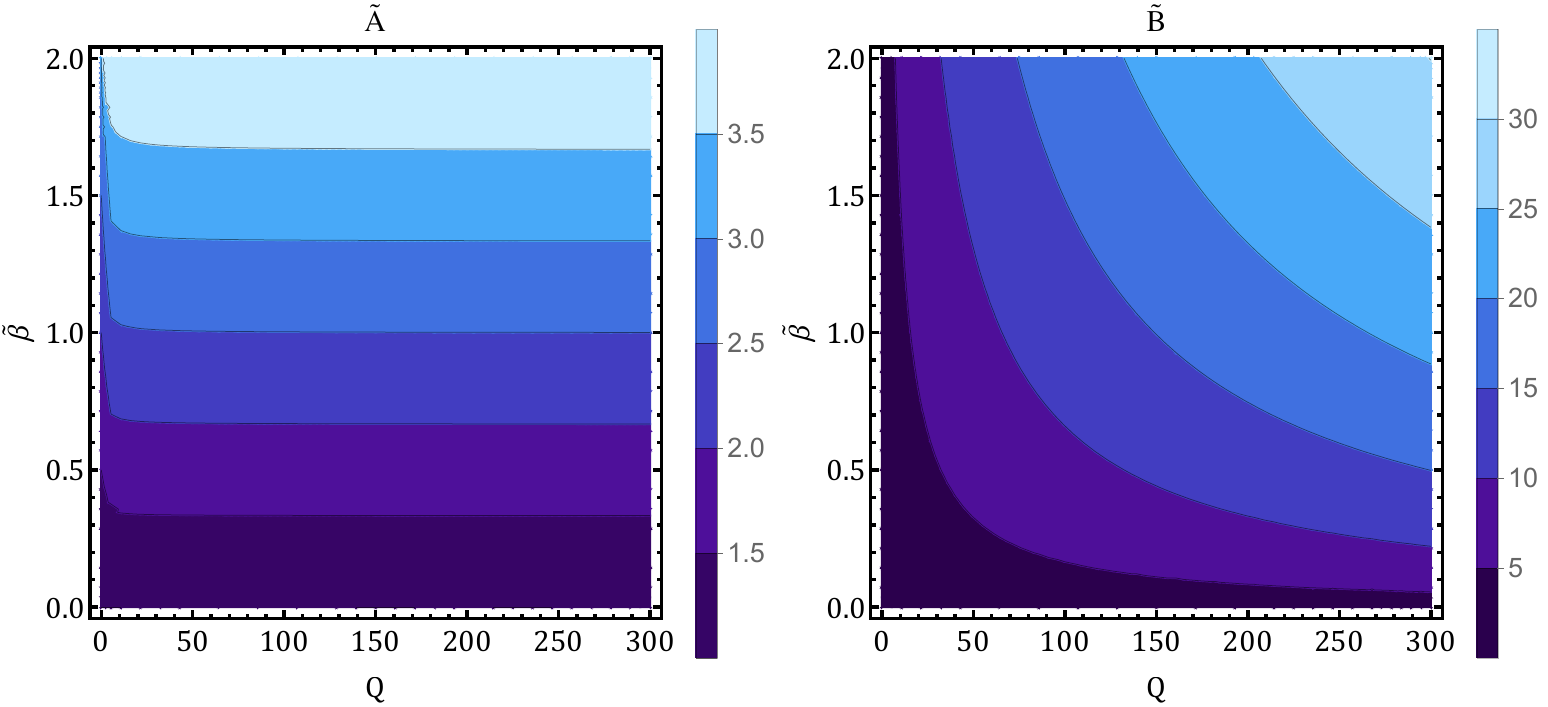}
	\caption{The variation of \(\mathcal{\Tilde{A}}\) and $\mathcal{\Tilde{B}}$ in the parameter space of \((Q, \Tilde{\beta})\) when $\beta<0$. }
	\label{fig:a_plot_nb}
\end{figure*}

In this case mostly all of the equations used to study the dynamics of the inflationary system remains the same as in the previous case except the constnt-roll condition. When $\beta<0$, the constraint equation for constant-roll can be written as:
\begin{equation}
	\ddot{\phi}=3\Tilde{\beta}H\dot{\phi}\label{constraint_eq_nb},
\end{equation}
where $\Tilde{\beta}=-\beta$.
When the constant roll parameter $\beta<0$, and during inflation the radiation energy density evolves slowly, constant-roll warm inflation is possible  with the inflaton potential of the form described in Ref.~\cite{Biswas:2024oje}:
\begin{equation}
	V(\phi)=V_{0}\left[1-\mathcal{\Tilde{A}}\sin^2{\left(\mathcal{\Tilde{B}}\kappa(\phi_{0}+\phi)\right)}\right]\label{v_crwi_neg_b}\,,
\end{equation}
where  $\mathcal{\Tilde{A}}\equiv \dfrac{2\left(1+Q\right)+2\Tilde{\beta}+3\Tilde{\beta} Q}{2(1+Q)}$ and $\mathcal{\Tilde{B}}\equiv 3\sqrt{\frac{(1+Q)\Tilde{\beta}}{6}}$ and $Q$ is a constant. Notice that unlike the $\beta>0$ case here $\mathcal{\Tilde{A}}$ and $\mathcal{\Tilde{B}}$ are always positive quantities. The variation of $\mathcal{\Tilde{A}}$ and $\mathcal{\Tilde{B}}$ with $\Tilde{\beta}$ and $Q$ is shown in Fig. [\ref{fig:a_plot_nb}]. As in the previous case, in the present case also the potential $V(\phi)$ is not positive definite. For our study of inflationary dynamics we will work with specific parameter values and some class of initial conditions which  always keeps the potential positive.

As established earlier, we adopt the dynamical variables \((x, y, z)\), \((\Omega_r, \Omega_\phi)\), and their compactified counterparts \((\bar{x}, \bar{z})\) as defined in Eqs.~(\ref{var_Crwi})–(\ref{compactification1}). These definitions ensure a finite and well-behaved phase space, mapping infinities to the interval \([0, 1)\), and will be used throughout the subsequent analysis. As in the previous case, we will only work with the positive branch of $x$.\\
The field $\phi$ can be expressed in terms of $\bar{x}$ as:
\begin{align}   
	\frac{1}{\kappa\mathcal{\Tilde{B}}}\arcsin\left(\sqrt{\frac{1}{\mathcal{\Tilde{A}}}\dfrac{1-4\bar{x}^2}{1-\bar{x}^2}}\right)-\phi_0&=\phi\label{crwi_phixbar_nb}\,.
\end{align}
This condition imposes a constraint on $\bar{x}$, specifically  $|\bar{x}|\leq \frac{1}{2}$. 
Here Eq. (\ref{crwi_phixbar_nb}) and Eq. (\ref{v_crwi_neg_b}) jointly yield:
\begin{align}
	V_{,\phi}=2 V_{0}\kappa\mathcal{\Tilde{A}}\mathcal{\Tilde{B}}\sqrt{\frac{1}{\mathcal{\Tilde{A}}}\dfrac{1-4\bar{x}^2}{1-\bar{x}^2}}\sqrt{1-\frac{1}{\mathcal{\Tilde{A}}}\dfrac{1-4\bar{x}^2}{1-\bar{x}^2}}\label{vphi_crwi_nb}\,.
\end{align}
The field equation for $\beta<0$ is:
\begin{align}
	3H\dot{\phi}(\Tilde{\beta}+1+Q)=-V_{,\phi},
\end{align}
which gives
\begin{equation}
	y= \pm \frac{2}{\sqrt{6}(\Tilde{\beta}+1+Q)}\frac{\sqrt{1-\zb^2}}{\zb}\mathcal{\Tilde{A}}\mathcal{\Tilde{B}}\sqrt{\frac{1}{\mathcal{\Tilde{A}}}\dfrac{1-4\bar{x}^2}{1-\bar{x}^2}}\sqrt{1-\frac{1}{\mathcal{\Tilde{A}}}\dfrac{1-4\bar{x}^2}{1-\bar{x}^2}}\label{y_crwi_nb}\,.
\end{equation}
Using Eq. (\ref{crwi_phixbar_nb}) and the definition of $\bar{z}$, we obtain the autonomous equations in the present case as:
\begin{eqnarray}
	\bar{x}'=&\sqrt{6}y \mathcal{\Tilde{A}}\mathcal{\Tilde{B}}\frac{(1-\bar{x}^2)^{2}}{3\xb}\sqrt{\dfrac{(4\xb^2-1)(1+\xb^2(4-\mathcal{\Tilde{A}})-\mathcal{\Tilde{A}})}{(1-\bar{x}^2)^{2}\mathcal{\Tilde{A}}^2}}\label{auto_eq_nbi}\,,\\
	\bar{z}'=&\left(1-\bar{z}^2\right)^{\frac{3}{2}}\left(-y^2-2\dfrac{\bar{z}^2}{1-\bar{z}^2}+2\dfrac{\bar{x}^2}{1-\bar{x}^2}\right)\label{auto_eq_nb1}\,,
\end{eqnarray}
where $y$ is given by Eq. (\ref{y_crwi_nb}), making the phase space two dimensional and where \(()' \equiv \frac{d()}{\kappa \sqrt{V_0} dt}\).  
We can see that $y$ has two branches, the positive and the negative ones. Here we have worked with the positive branch to illustrate the results.  For warm inflation we need the radiation energy density $\rho_{r}>0$, i.e.
\begin{equation}
	\frac{\rho_r}{3V_0}=\dfrac{\bar{z}^2}{1-\bar{z}^2}-\dfrac{\bar{y}^2}{1-\bar{y}^2}-\dfrac{\bar{x}^2}{1-\bar{x}^2}>0\,.
\end{equation}
During the inflationary  phase we need $\epsilon_{1}<1$ and we demand that inflation occurs near a dynamic thermal equilibrium, which translates to the condition: $ \frac{|\dot{\rho}_r|}{4H\rho_r}<1$.

To analyze the behavior of the system, we utilize the autonomous system of equations previously constructed. The autonomous equations corresponding to the variables \((\bar{x}, \bar{z})\) are given by Eq.~(\ref{auto_eq_nbi}) and Eq. \eqref{auto_eq_nb1}. In the present case the autonomous system of equations diverges for \(\bar{x} = 0\) and \(\bar{z} = 0\). To address these divergences, we redefine the time variable as follows:   
\begin{equation}
	\kappa \sqrt{V_0} dt  \to \kappa \sqrt{V_0} \bar{z}^2 \bar{x} d {t} \,. 
\end{equation}

The modified autonomous equations are:
\begin{eqnarray}
	\bar{x}'= &\sqrt{6}y \mathcal{\Tilde{A}}\mathcal{\Tilde{B}}\frac{\bar{z}^{2}(1-\bar{x}^2)^{2}}{3}\sqrt{\dfrac{(4\xb^2-1)(1+\xb^2(4-\mathcal{\Tilde{A}})-\mathcal{\Tilde{A}})}{(1-\bar{x}^2)^{2}\mathcal{\Tilde{A}}^2}}\,,\\
	\bar{z}'= &\bar{z}^{2}\bar{x}\left(1-\bar{z}^2\right)^{\frac{3}{2}}\left(-y^2-2\dfrac{\bar{z}^2}{1-\bar{z}^2}+2\dfrac{\bar{x}^2}{1-\bar{x}^2}\right)\,.
\end{eqnarray}
\begin{table}[t]
	\centering
	\begin{tabular}{|c|c|c|c|c|c|}\hline
		
		Points&  \((\bar{x}_*, \bar{z}_*) \)&  \(\Omega_\phi \)&  \(\epsilon_1 \)& \(\frac{\dot{\rho}_{r}}{H \rho_{r}}\) & Stability \\ \hline 
		
		$P_{1}$ & $(0,0)$ & $\infty$ & $\infty$ & $-\frac{3 Q}{2}-1$ & Unstable\\

		\hline
		
		%
		
		$P_{2}$ & $(\rm{Any}, 1)$ & $0$ & $2$ & $-1$ & Unstable\\
		\hline
		
	\end{tabular}
	\caption{The critical points and their nature for CRWI Model. }
	\label{tab:warm_constant_nb}
\end{table}
To determine the qualitative behavior of the dynamical system we compute the critical points corresponding to the redefined equations. These critical points are summarized in Tab.~[\ref{tab:warm_constant_nb}]. The system yields two critical points, none of which depend on any model parameters, including \(\tilde{\beta}\). At point \(P_1\), both coordinates vanish, resulting in an indeterminate inflaton fractional energy density. Additionally, the slow-roll parameter becomes very large, rendering this point unstable.
%


The points \(P_2\) actually specifies a line of fixed points and these fixed points correspond to phases in the radiation dominated epoch, as \(\Omega_\phi\) vanishes and the slow-roll parameter exceeds 1. Upon evaluating their stability, both the points \(P_{1}\) and \(P_2\) exhibit unstable nature for any choice of model parameters. 

\begin{figure}[t]
	\centering
	\includegraphics[scale=0.24]{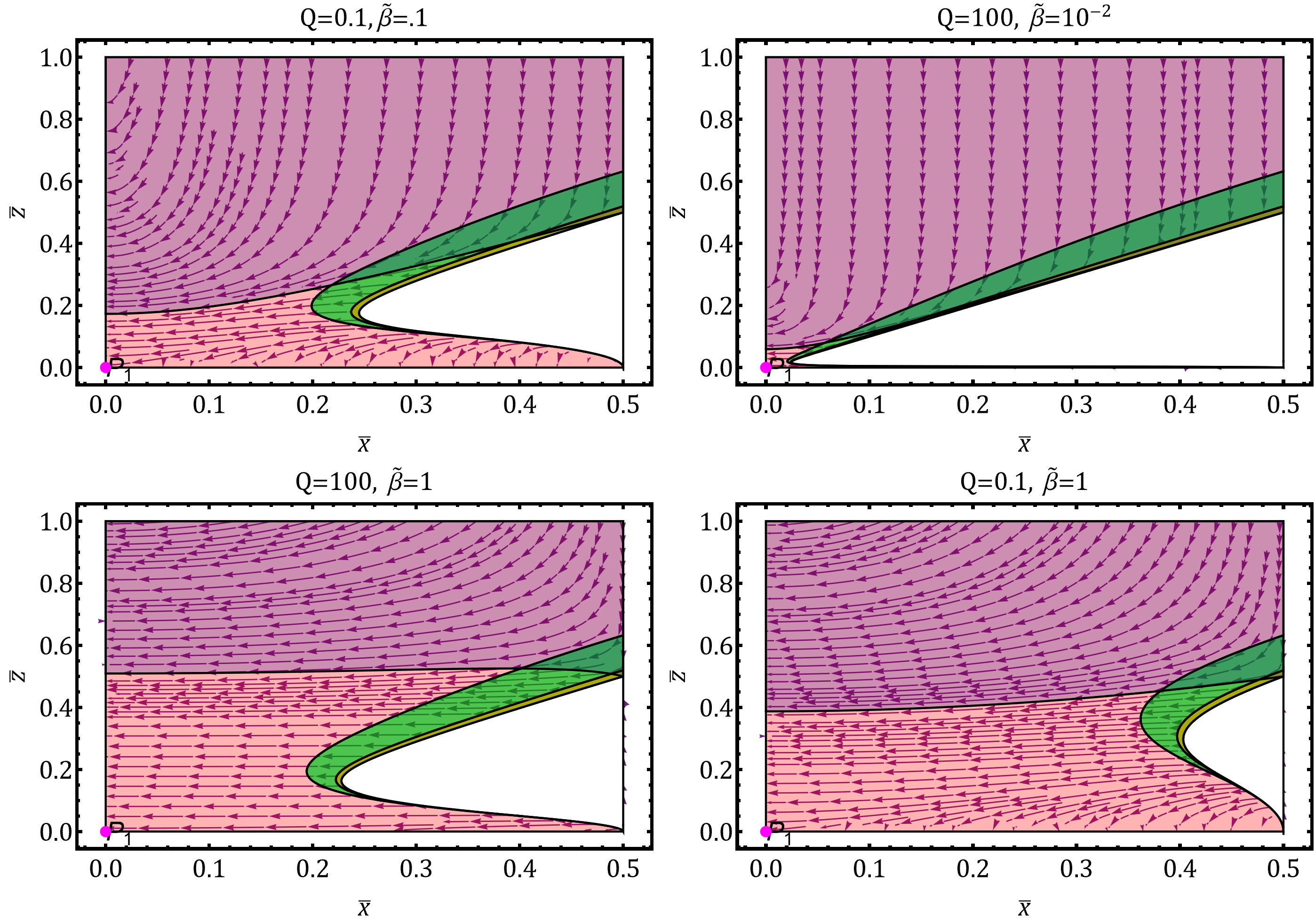}
	\caption{The phase space corresponding to CRWI with \(\beta<0\). The color scheme highlights the different dynamical regions: pink (\(\epsilon_1 \geq 1\)), yellow region  (\(0 \leq \epsilon_1 \leq 0.1\)), green region  (\(0.1 < \epsilon_1 < 1\)), blue (thermally stable region), and violet (overlap of red and blue regions).}
	\label{phase_crwi_tilde}
\end{figure}

The phase space for the constant-roll warm inflation (CRWI) model is illustrated in Fig.~[\ref{phase_crwi_tilde}] for various parameter choices of \((Q, \tilde{\beta})\) within the \((\bar{x}, \bar{z})\) space. The phase space is depicted using distinct color codes: the pink region corresponds to \(\epsilon_1 \geq 1\), the yellow region corresponds to \(0 \leq \epsilon_1 \leq 0.1\), the green region corresponds to \(0.1 < \epsilon_1 < 1\)  and finally the blue region specifies the set of points in the phase space where we have thermal stability . Where the blue and red regions overlap, the color appears violet. Additionally, the white region represents areas where the slow-roll parameter is negative (\(\epsilon_1 < 0\)) and hence these regions are practically ruled out from any meaningful dynamical analysis. Since \(\bar{z} = 1\) corresponds to a line of unstable fixed points, Fig.~[\ref{phase_crwi_tilde}] illustrates that trajectories diverge away from it. Additionally, the figure shows that all trajectories move toward \( V = 0 \), i.e., the \( \bar{x} = 0 \) line.

The plots on the lower panels show that when \(\tilde{\beta} \gtrsim 1\), the thermally stable blue region has little to no overlap with the green or yellow regions, making CRWI practically impossible in these cases. In contrast, for the figures in the upper panels where \(0 < \tilde{\beta} \ll 1\), a substantial overlap between the blue and yellow regions ensures the existence of a thermally stable inflationary phase.  In this favorable regime, the system's trajectories originate from the critical points at \(P_2\) (corresponding to the \(\bar{z} = 1\) line), traverses the inflationary epoch (highlighted in yellow), and eventually transitions into the inflationary region (green) while maintaining thermal stability at the radiation-dominated phase. This behavior closely resembles the dynamics observed for \(\beta>0\), reinforcing the underlying qualitative consistency across these models.

\begin{figure}[t!]
	\centering
	\includegraphics[scale=0.5]{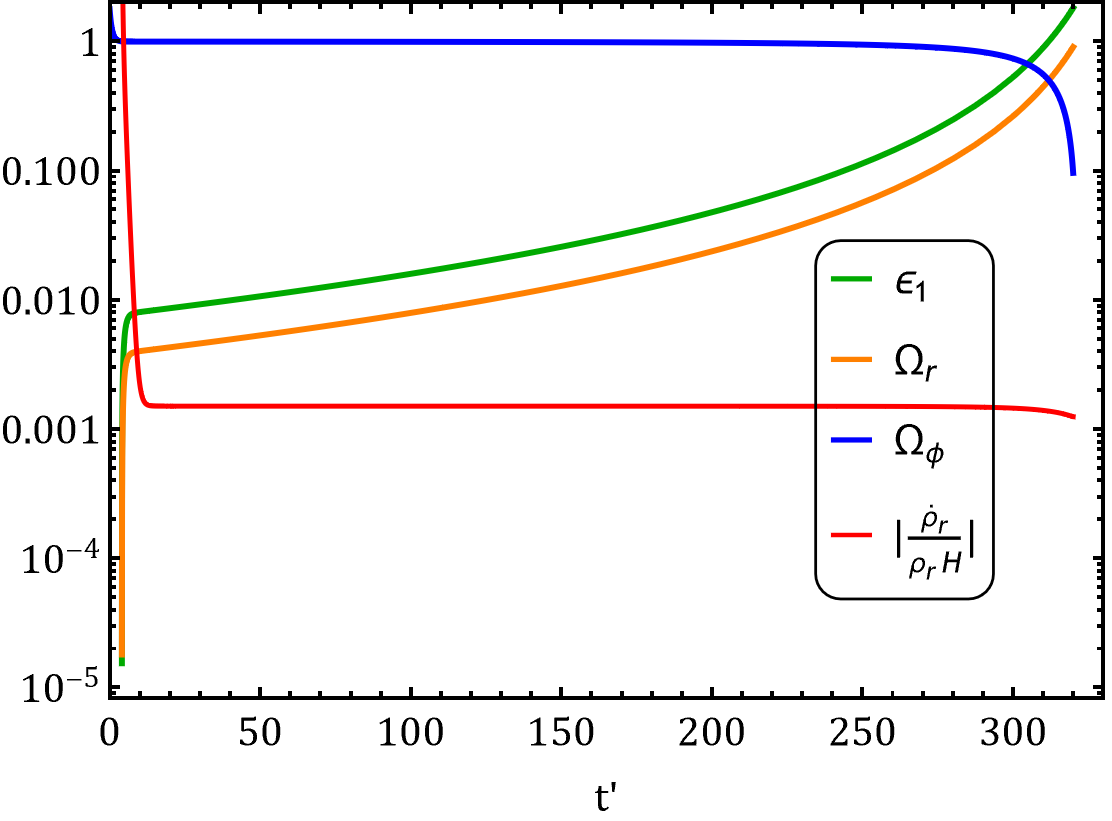}
	\includegraphics[scale=0.8]{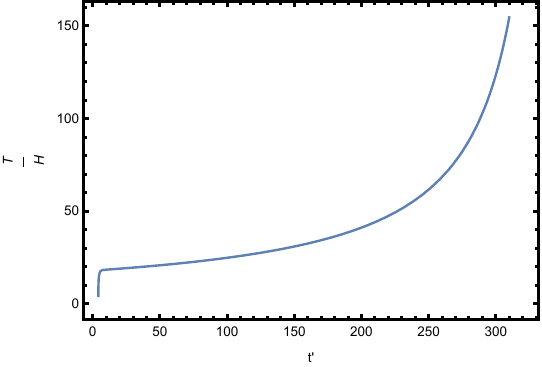}
	\caption{The dynamical evolution of various variables in CRWI and the evolution of $T/H$  during inflation corresponding to $(Q=300, \tilde{\beta} = 10^{-3})$ and $V_{0}=10^{-4} M^{4}_{\rm Pl}$.}
	\label{cosmo_params_tilde}
\end{figure}

We can choose one of the phase space trajectories to roughly figure out the inflationary dynamics in the present case. The temporal evolution of the cosmological parameters is shown in left hand panel of Fig.~[\ref{cosmo_params_tilde}] for a representative parameter set \((Q = 300, \tilde{\beta} = 10^{-3})\). Initially, the field energy density dominates over the radiation energy density, resulting in the slow-roll parameter \(\epsilon_1 < 1\), which confirms the system is within the inflationary phase.  This plot shows a rough estimate of the inflationary process because we have shown the dynamics corresponding to a trajectory in the phase space and in this case we have not concentrated explicitly on the parameter choice and exact initial conditions which can produce an ideal inflationary phase. The right hand panel of  Fig.~[\ref{cosmo_params_tilde}] shows that throughout the inflationary phase we have $T>H$.

As the radiation energy density begins to dominate, the slow-roll parameter exceeds 1, marking the system’s exit from the inflationary regime after \(t'=320\) which corresponds to about 56 $e$-folds.  A more robust choice of parameters  and initial conditions may have increased the number of $e$-folds, but in this paper we do not specifically concentrate on the question of tuning the initial conditions and parameters. Throughout the evolution, the modulus of the thermal stability parameter remains less than 1,  ensuring thermal stability is maintained during the entire process. This behavior is consistent with the corresponding analysis for \(\beta>0\) case, emphasizing the robustness of the thermally stable inflationary dynamics in both parameterizations ($\beta \lessgtr 0$). 

Comparing the phase space plots obtained in the various cases, in the present circumstance, with those obtained in the case of CRCI and WI, we see many differences. If we compare with the warm inflation case we see that unlike a cylindrical phase space we have a two dimensional phase space. Using compactified variables this two dimensional phase space becomes becomes closed and bounded. We see that in the present case a large part of the region inside the probable phase space becomes unphysical, as in those regions one gets a negative value of the slow-roll parameter $\epsilon_1$. The phase plots display a reduced number of trajectories which gives rise to a thermalized CRWI with graceful exit. For some parameter values we do not obtain any inflationary orbits. If we compare the phase orbits obtained in the present section with those obtained in the case of CRCI then we see CRWI is more restrictive. These feature is expected and apparent from the phase space behavior of the dynamical systems. 

\section{Warm ultraslow-roll (USR)  inflation and its stability}
\label{wusr}

In warm inflation, an ultraslow-roll (USR) phase refers to a period during which the inflaton rolls on an almost flat section of the potential, but unlike standard slow-roll, the acceleration term \(\ddot{\phi}\) is not negligible. In fact, it plays a crucial role in the dynamics. The inflaton equation of motion in warm inflation takes the form
\begin{equation}
	\ddot{\phi} + (3H + \Upsilon)\dot{\phi} + V_{,\phi} = 0,
\end{equation}
where \(\Upsilon\) is the dissipation coefficient. During USR, the potential gradient \(V_{,\phi} \equiv dV/d\phi\) is much smaller than the friction and the acceleration term, so the field evolution is driven mainly by the damping terms and its own acceleration, i.e., $\ddot{\phi} + (3H + \Upsilon)\dot{\phi} = 0$. Compared to the cold inflation case, the presence of \(\Upsilon\) adds an extra source of friction, which can affect both the duration and stability of the USR phase. Depending on the strength of dissipation, the system may linger in USR longer, or exit it more quickly. Understanding USR in warm inflation is especially interesting since it can affect the generation of curvature perturbations, and may have implications for features like enhanced power spectra or even primordial black holes\cite{Dimopoulos_2017, Pattison_2018, Mishra_2020}. 

\subsection{Dynamical analysis of warm USR inflation in the potential $V(\phi)=V_{0}+M^3 \phi$}

In Ref. \cite{Biswas_2024}, the authors explored the scope of warm USR inflation with a linear potential of the form:
\begin{equation}
	V(\phi)=V_{0}+M^3 \phi \ ,\label{usr_lin_v}
\end{equation}
where $V_{0}$ is a density constant, and $M$ is a mass scale. Additionally, it has been assumed that $\kappa V_{0}\gg M^3$ consequently $ V_{0}\gg M^3\phi$. Under this condition, the dynamics of USR has been studied. \\
In this section, we explore the dynamics of the present scenario using the dynamical systems stability framework. The presence of higher-derivative terms in the current model may lead to novel dynamical features that have not been addressed in previous studies.
Following the same approach as before, we define the dimensionless variables:
\begin{align}
	x\equiv\sqrt{\frac{V(\phi)}{3V_{0}}}, & & y\equiv{\frac{\dot{\phi}}{\sqrt{6 V_{0}}}},& & z\equiv\frac{H}{\kappa\sqrt{V_{0}}}\label{usr_var} \ .
\end{align}
Ref. \cite{Biswas_2024} demonstrated that achieving ultra-slow-roll (USR) inflation within the warm inflation (WI) framework requires the dissipation coefficient $\Upsilon$ to depend, at minimum, on the inflaton field $\phi$, i.e., $\Upsilon = \Upsilon(\phi, T)$. In particular, Ref. \cite{Biswas_2024} considered the form $\Upsilon(\phi, T) = C_{\Upsilon} \frac{T^3}{\phi^2}$, where $C_{\Upsilon}$ is a dimensionless constant. The equations governing the WI in this scenario are:
\begin{align}
	\ddot{\phi}+3H\dot{\phi}+C_{\Upsilon}\frac{T^{3}}{\phi^{2}}\dot{\phi}+M^{3}&=0 \ ,\label{usr_wi_phi_eq}\\
	\dot{\rho}_{r}+4H\rho_{r}&=C_{\Upsilon}\frac{T^{3}}{\phi^{2}}\dot{\phi}^2\label{usr_wi_rho_eq} \ .
\end{align}
Using the Friedmann equation (Eq.~\ref{first_fried_wi}), we can express the radiation energy density in terms of our dimensionless variables:
\begin{equation}
	\frac{\rho_r}{3V_0} = z^2 - y^2 - x^2.
\end{equation}
Now we construct the autonomous system of equations in terms of the dimensionless dynamical variables using the field equation as: 
\begin{eqnarray}
	& & x' =\frac{M^{3}\sqrt{6}}{\kappa V_0}\frac{y^2}{x} \ , \label{w_usr_xeq} \\
	&&  y' = -\left(  3yz + C_{\Upsilon}C_R^{3/4}27^{1/4}M^6\frac{(z^2 - y^2 - x^2)^{3/4}}{\kappa V_0^{7/4}(3x^2 - 1)} + \frac{M^3}{\kappa V_0} \right) \ , \label{usr_wi_y_eq} \\
	&&  z' = 2x^2 - y^2 - 2z^2 \ .
\end{eqnarray}
where $()'\equiv\frac{1}{\kappa \sqrt{V_{0}}}\frac{d}{dt}$ and we define $t'=\kappa \sqrt{V_{0}}t$. For this scenario, the phase space becomes 3-dimensional, however,the system does not have any critical points. Hence, we will conduct our further analysis numerically. Before we do, at first we express few important quantities in terms of dynamical variables. For instance, the Hubble slow roll parameter $\epsilon_{1} \equiv  -\frac{\dot{H}}{H^{2}} $ can be expressed in terms of dynamical variables as:
\begin{align}
	\epsilon_{1}&=\frac{2x^2 - y^2 - 2z^2}{z^2} \ , 
\end{align}
and similarly, the second slow roll parameter can be expressed as 
\begin{equation}
	\epsilon_{2} \equiv \frac{\dot{\epsilon_{1}}}{H\epsilon_{1}} =2\epsilon_{1}+\frac{4xx'-2yy'-4zz'}{z(2x^2 - y^2 - 2z^2)} \ ,
\end{equation}

\begin{figure}
	\centering
	\includegraphics[scale=0.35]{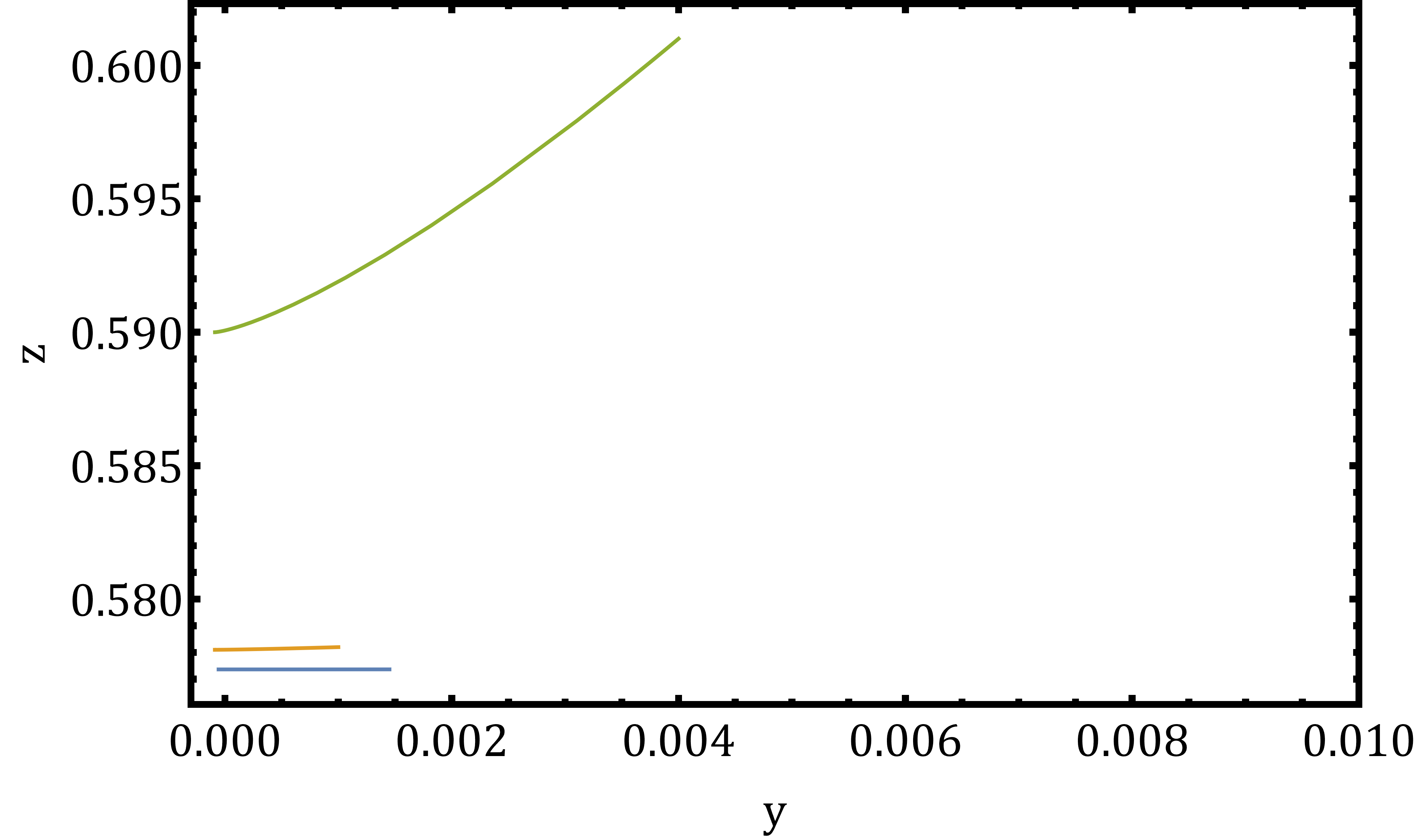}
	\caption{Phase-space dynamics in the $y$-$z$ plane showing {orange}, 
		{blue}, and {green} trajectories corresponding 
		to three different initial conditions. The colored paths 
		reveals the non-attractor behavior intrinsic to USR inflation.}
	\label{fig:ps_usr}
\end{figure}

The {ultra-slow-roll (USR) phase} is characterized by the conditions $\epsilon_1 < 1$ but $|\epsilon_2| > 1$. In case of WI, the system must satisfy $|\frac{1}{H}\frac{\dot{\rho}_{r}}{\rho_{r}}|<1$. Once the USR phase ends after a few e-folds, the inflaton field transition to the {slow-roll (SR) phase}, marked by $\epsilon_1 < 1$ and $|\epsilon_2| < 1$.
We use a plot showing the variation of  \(y\text{--}z\) in Fig.~[\ref{fig:ps_usr}], because the full three-dimensional trajectory does not provide clear information about whether the system converges to any particular region in phase space. In contrast, the 2D parametric plot offers more insight into the USR phase and other dynamical regimes. However, the curves in this plot do not appear to converge to any particular region in the phase space, indicating that the USR phase is not a stable attractor.
In the parametric plot, we show three distinct trajectories corresponding to three different sets of initial conditions. These are represented by the {orange}, {blue}, and {green} curves, each following noticeably different evolutionary paths during the USR regime. The parameters used are \(V_{0} = (10^{-4})^4 M^4_{\mathrm{pl}}\), \(M = 2.5 \times 10^{-7} M_{\mathrm{pl}}\), \(C_{\Upsilon} = 10^5\), and \(g_{*} = 106.75\). The {orange} trajectory starts from \(x(0) = 5.77365 \times 10^{-1}\), \(y(0) = 1.4506 \times 10^{-3}\), \(z(0) = 5.77366 \times 10^{-1}\); the {blue} trajectory begins at \(x(0) = 5.781 \times 10^{-1}\), \(y(0) = 1 \times 10^{-3}\), \(z(0) = 5.782 \times 10^{-1}\); and the {green} trajectory originates from \(x(0) = 5.9 \times 10^{-1}\), \(y(0) = 4 \times 10^{-3}\), \(z(0) = 6.01 \times 10^{-1}\). As time evolves, all three trajectories moves leftward in the \(y\text{--}z\) plane, yet none of them converge to a common point, reinforcing the interpretation that the USR phase does not correspond to a stable solution. Moreover, we observe only portions of these trajectories in the \(y\text{--}z\) plane because the system of differential equations becomes stiff after a certain time, making numerical integration impossible beyond that point.

To study the initial condition of the system and the fate of warm USR it is more suitable if we follow the approach of 
Ref.~\cite{Pattison_2018}. In warm inflationary scenario we can generalize the treatment in \cite{Pattison_2018} and  define the dimensionless parameters:
\begin{align}
	f_{\phi}=-\frac{\ddot{\phi}}{3H\dot{\phi}+\Upsilon \dot{\phi}}, && f_{\rho}=\frac{\dot{\rho}}{4H\rho}\ .
\end{align}
In Warm Ultra-Slow-Roll (USR) regime, these parameters exhibit $f_{\phi} \simeq 1$,and $|f_{\rho}| < 1$, reflecting production of radiation energy maintaining thermal equilibrium. These parameters distinguishes between different inflationary regimes: (i) \emph{slow-roll} (\text{SR}) regime: $|f_{\phi}| \ll 1$ and (ii) \emph{ultra-slow-roll} (\text{USR}) regime: $|f_{\phi} - 1| \ll 1$, while maintaining $|f_{\rho}| < 1$ through out the evolution in both cases. We linearise $f$ about $f=1$ by parameterising 
\begin{equation}
	f_{\phi}=1-\delta\ ,
\end{equation}
where $|\delta|\ll 1$. As $|\delta|$ increases (decreases) with time the USR becomes unstable (stable). As the predefined variables can take any arbitrary values, now we compactify those using the Poincar\'e transformation 
\begin{align}
	&x\to \frac{x}{\sqrt{1+x^2}}, &y\to\frac{y}{\sqrt{1+y^2}},& &z\to\frac{z}{\sqrt{1+z^2}},
\end{align}
resulting in $(x,y)\in [-1,1]$ and $z \in [0,1]$ range. Compactifying the variables helps us to choose the initial conditions as now we can in principle choose any allowable points in the compactified phase space as initial conditions. If we have not compactified the phase space, points at infinity or points far away from the origin may have been left out.  

For the given potential in Eq.~(\ref{usr_lin_v}), the system cannot start in the SR regime where \(\delta\sim 1\). Because, if we choose the initial conditions corresponding to such cases, the system gets stuck in the SR regime which is the attractor phase and unable to render the USR phase. This is our main prediction about the initial condition in warm USR inflation. In a certain sense warm USR is very different from other models of inflation as here one cannot have a transient warm USR phase terminating in a SR phase unless one starts with initial conditions which do satisfy USR conditions, hence we may infer that such a transcient phase lacks an attractor solution as general initial conditions are not attracted towards warm USR. 
Hence, we choose the initial conditions corresponding to the USR phase i.e., $\delta \ll 1$. 
Now we plot $f_{\phi}$ for the transformed variables in Fig.~[\ref{fig:fphi_lin}]. The results indicate optimal behavior when the initial values are given as $x_0 \approx 0.51$, with $z_0 \gtrsim x_0$ and $y_0 \ll |x_0|$. As we increase $x$ beyond this point, the system spends progressively less time in the USR phase, marked by $f_\phi \sim 1$ and transit to $f_\phi \to 0$, SR phase. For $x_0 < 0.51$, $f_{\rho} <- 1$, which violates the energy conditions.\\
\begin{figure}[htbp]
	\centering
	\includegraphics[scale= 0.32]{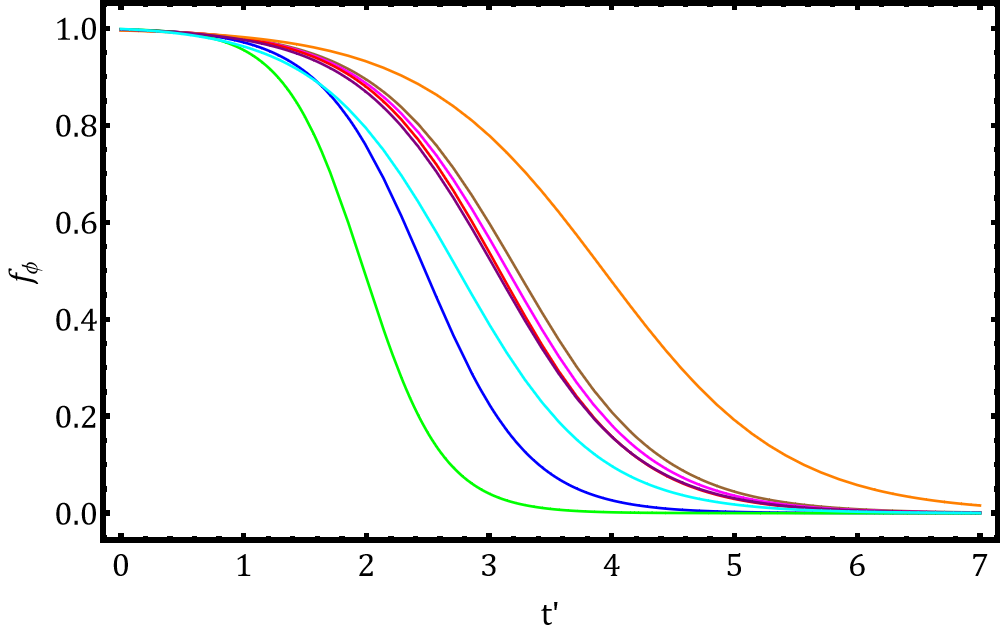}
	\includegraphics[scale=0.35]{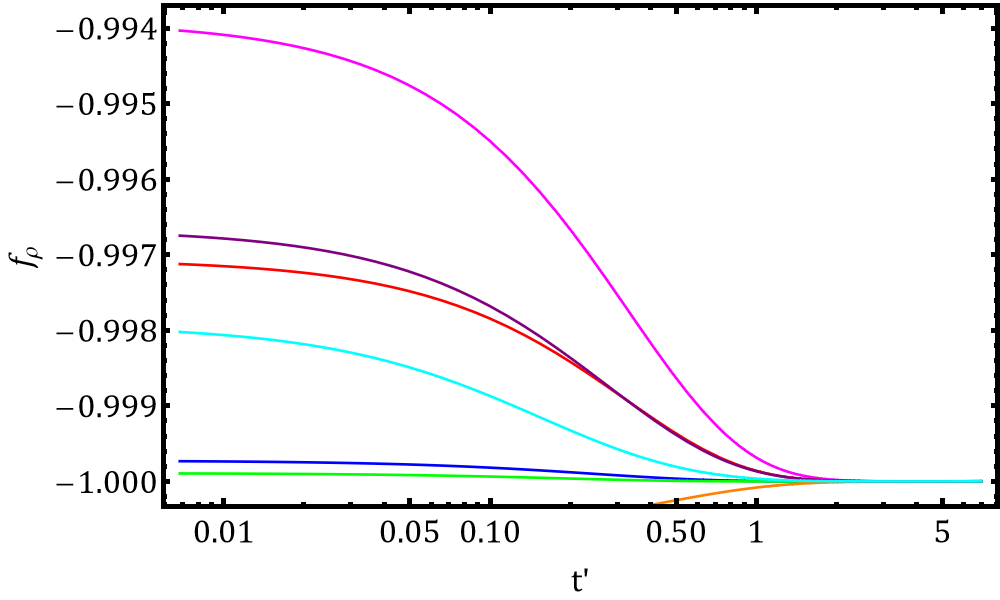}
	\caption{
		Evolution of the dissipation coefficient $f_\phi, f_\rho$ for a linear potential $V(\phi) = V_0 + M^3\phi$. 
		The limit $f_\phi \to 0$ signals the onset of slow-roll inflation. Also $f_{\rho} >- 1$ indicates that the system in those cases evolves maintaining thermal equilibrium, while for the orange line, which corresponds to $x_0 < 0.51$, we have $f_{\rho} <- 1$, which violates the condition $|f_\rho|<1$ necessary for thermalization. 
	}
	\label{fig:fphi_lin}
	\label{fig:fphi_lin}
\end{figure}

To visualize the impact of initial conditions, we have plotted several distinct curves in the 2D plots in Fig.~[\ref{fig:fphi_lin}], each corresponding to a different set of initial values of \((x_{0}, y_{0}, z_{0})\), and each assigned a unique color. The initial conditions span the ranges \(x_{0} \in [0.4, 0.75]\), \(y_{0} \in [-10^{-3}, -10^{-2}]\), and \(z_{0} \in [0.4, 0.8]\). The {brown}, {magenta}, and {red} curves represent variations around a fiducial trajectory with small deviations in all three variables. The {blue} and {green} curves correspond to larger values of \(x_{0}\), with \(y_{0}\) and \(z_{0}\) held near their fiducial values. The {purple} and {cyan} curves explore significant deviations in \(z_{0}\), while keeping \(x_{0}\) and \(y_{0}\) fixed. The {orange} curve illustrates the effect of decreasing \(x_{0}\), with \(y_{0}\) and \(z_{0}\) again held fixed. This color scheme allows us to track how variations in each parameter influence the evolution of \(f_\phi\), particularly in identifying conditions that either extend the USR phase or lead to an earlier transition into the slow-roll regime. We observe that in all cases, the system eventually transitions toward the slow-roll attractor, indicating that the USR phase is a transient phase.

{Previous authors \cite{Passaglia:2018ixg, Cai:2018dkf} have noted that cold USR phase is a non-attractor phase. In this work we have shown that even warm USR inflation is not the standard attractor solution as there may be initial states in the phase space (which do not obey the USR condition) which may not ultimately stabilize to a slow-roll phase after the USR phase. One may say the warm USR inflation is a particular form of non-attractor solution with more subtlety. The subtlety arises because one has to keep in account the thermalization condition. In this section it is  specifically shown which kind of initial conditions can give rise to a warm USR phase followed by a slow-roll phase maintaining the thermalization condition. If we deviate from the thermalization condition the stability analysis will get modified. Our work shows the way in which the attractor solution fails in the present case and opines about the stability of warm USR phase of inflation.}

\section{Conclusion \label{sec:conclusion}}

In this work we have tried to formulated a methodical approach to study the qualitative as well as the quantitative features of the phase space of the nonstandard inflationary models. We have studied non-slow-roll inflationary models where one of the models do not have any slow-roll limit. Our primary emphasize has been on the possible initial conditions which can give rise to various non-slow-roll inflationary phases.

Our work presents some new results. Some of these new results are general and some are particular. In this paper we have for the first time shown that there can be a particular potential, as discussed in section \ref{crci}, in which one can have CRCI. This result is new as previously the authors in Ref.~\cite{Motohashi_2015} the authors opined that CRCI cannot occur in the aforementioned potential. We have specified the nature of initial conditions in CRCI for various values of the CRCI parameter $\beta$. Next we have for the first time discussed the phase space description of CRWI.

Compared to standard cold inflation and warm inflation models, the constant-roll variants of these models are relatively constrained. The constraint comes from the constant-roll condition and is dynamic in nature. This dynamic constraint has appreciable effect on the dynamics as it drastically affects the dimension as well as the morphology of the phase space. 
The constant-roll condition directly affects the issue of graceful exit from inflation. Moreover the constant-roll condition transforms the dynamical system of CI into an algebraic system where one can figure out the phase space dynamics without explicitly solving any set of autonomous equations. Due to the constant-roll condition, the actual phase space of CRCI is one dimensional. To make the results more tractable we have plotted a recreated two dimensional stream-plot where the streamlines have phase space coordinateds in the case of CRCI. Compared to CRCI, CRWI has a genuine two dimensional phase space and CRWI can be tackled both algebraically as well as with the system of autonomous differential equations. In the present paper we have chosen the system of autonomous equations to study the dynamics of CRWI.  

In the last section we have extended our approach to the case of warm USR inflation. In the USR variation of WI, we have shown that the system does not admit general initial conditions. Only those initial conditions are allowed here, for which we alrady have the USR condition fulfilled. Only these initial states can ultimately have a transient USR inflation phase which terminates in a SR inflation phase. From these findings it can be seen that entering an USR phase in WI is in general difficult and not this phase requires very particular kind of initial conditions. These results are obtained using a particular form of the inflaton potential, but we expect that the results are fairly general and will hold true for other kind of inflaton potentials giving rise to a transient USR phase.   

Inflationary phenomenology is in general aimed at producing 60-70 $e$-folds of inflation and thereby producing cosmological perturbations so that structures can be formed in the later part of cosmological evolution. More emphasize is given to the calculation of power spectrum which is generated from various inflationary models and comparing those results with observational data. The  dynamical stability and attractor nature of the background model in inflation is studied rarely.  This study is very important as because the inevitability of cosmological inflation greatly depends upon the attractor nature of the inflationary processes in the phase space. In this paper we have introduced a thorough and methodical way to systematically study the dynamical structure and stability of the background inflationary models for non-slow-roll models of inflation. We hope in the near future we will also be able to propose a methodical way to study the growth of inflationary perturbations.

\section*{Acknowledgement}

S.H. acknowledges the support of National Natural Science Foundation of China under Grants No. W2433018 and No.
11675143, and the National Key Reserach and development Program of China under Grant No. 2020YFC2201503.\\
The authors acknowledge professor Suratna Das' comments regarding various points discussed in this paper.

\bibliographystyle{spphys}       
\bibliography{refe}   

@article{Kazanas:1980tx,
    author = "Kazanas, D.",
    title = "{Dynamics of the Universe and Spontaneous Symmetry Breaking}",
    doi = "10.1086/183361",
    journal = "Astrophys. J. Lett.",
    volume = "241",
    pages = "L59--L63",
    year = "1980"
}

@article{Guth:1980zm,
    author = "Guth, Alan H.",
    editor = "Fang, Li-Zhi and Ruffini, R.",
    title = "{The Inflationary Universe: A Possible Solution to the Horizon and Flatness Problems}",
    reportNumber = "SLAC-PUB-2576",
    doi = "10.1103/PhysRevD.23.347",
    journal = "Phys. Rev. D",
    volume = "23",
    pages = "347--356",
    year = "1981"
}

@article{Sato:1981ds,
    author = "Sato, K.",
    editor = "Fang, Li-Zhi and Ruffini, R.",
    title = "{Cosmological Baryon Number Domain Structure and the First Order Phase Transition of a Vacuum}",
    doi = "10.1016/0370-2693(81)90805-4",
    journal = "Phys. Lett. B",
    volume = "99",
    pages = "66--70",
    year = "1981"
}

@article{Sato:1980yn,
    author = "Sato, K.",
    title = "{First Order Phase Transition of a Vacuum and Expansion of the Universe}",
    reportNumber = "NORDITA-80-29",
    journal = "Mon. Not. Roy. Astron. Soc.",
    volume = "195",
    pages = "467--479",
    year = "1981"
}

@article{Linde:1981mu,
    author = "Linde, Andrei D.",
    editor = "Fang, Li-Zhi and Ruffini, R.",
    title = "{A New Inflationary Universe Scenario: A Possible Solution of the Horizon, Flatness, Homogeneity, Isotropy and Primordial Monopole Problems}",
    reportNumber = "LEBEDEV-81-229",
    doi = "10.1016/0370-2693(82)91219-9",
    journal = "Phys. Lett. B",
    volume = "108",
    pages = "389--393",
    year = "1982"
}

@article{Carrasco_2015,
   title={Cosmological attractors and initial conditions for inflation},
   volume={92},
   ISSN={1550-2368},
   url={http://dx.doi.org/10.1103/PhysRevD.92.063519},
   DOI={10.1103/physrevd.92.063519},
   number={6},
   journal={Physical Review D},
   publisher={American Physical Society (APS)},
   author={Carrasco, John Joseph M. and Kallosh, Renata and Linde, Andrei},
   year={2015},
   month=sep }

@article{Alho:2023pkl,
	author = "Alho, Artur and Uggla, Claes",
	title = "{Quintessential \ensuremath{\alpha}-attractor inflation: a dynamical systems analysis}",
	eprint = "2306.15326",
	archivePrefix = "arXiv",
	primaryClass = "gr-qc",
	doi = "10.1088/1475-7516/2023/11/083",
	journal = "JCAP",
	volume = "11",
	pages = "083",
	year = "2023"
}

@article{Albrecht:1982wi,
    author = "Albrecht, Andreas and Steinhardt, Paul J.",
    editor = "Fang, Li-Zhi and Ruffini, R.",
    title = "{Cosmology for Grand Unified Theories with Radiatively Induced Symmetry Breaking}",
    reportNumber = "UPR-0185T",
    doi = "10.1103/PhysRevLett.48.1220",
    journal = "Phys. Rev. Lett.",
    volume = "48",
    pages = "1220--1223",
    year = "1982"
}

@article{Riotto:2002yw,
    author = "Riotto, Antonio",
    editor = "Dvali, G. and Perez-Lorenzana, Abdel and Senjanovic, G. and Thompson, G. and Vissani, F.",
    title = "{Inflation and the theory of cosmological perturbations}",
    eprint = "hep-ph/0210162",
    archivePrefix = "arXiv",
    reportNumber = "DFPD-TH-02-22",
    journal = "ICTP Lect. Notes Ser.",
    volume = "14",
    pages = "317--413",
    year = "2003"
}

@article{Martin:2012pe,
    author = "Martin, Jerome and Motohashi, Hayato and Suyama, Teruaki",
    title = "{Ultra Slow-Roll Inflation and the non-Gaussianity Consistency Relation}",
    eprint = "1211.0083",
    archivePrefix = "arXiv",
    primaryClass = "astro-ph.CO",
    reportNumber = "RESCEU-47-12",
    doi = "10.1103/PhysRevD.87.023514",
    journal = "Phys. Rev. D",
    volume = "87",
    number = "2",
    pages = "023514",
    year = "2013"
}

@article{Biswas:2024oje,
    author = "Biswas, Sandip and Bhattacharya, Kaushik and Das, Suratna",
    title = "{Reassessing constant-roll warm inflation}",
    eprint = "2406.00340",
    archivePrefix = "arXiv",
    primaryClass = "astro-ph.CO",
    doi = "10.1103/PhysRevD.110.063536",
    journal = "Phys. Rev. D",
    volume = "110",
    number = "6",
    pages = "063536",
    year = "2024"
}

@article{Martin:2005ir,
	author = "Martin, Jerome and Musso, Marcello",
	title = "{Solving stochastic inflation for arbitrary potentials}",
	eprint = "hep-th/0511214",
	archivePrefix = "arXiv",
	reportNumber = "UTTG-14-05",
	doi = "10.1103/PhysRevD.73.043516",
	journal = "Phys. Rev. D",
	volume = "73",
	pages = "043516",
	year = "2006"
}

@article{ELIAS2006305,
	title = {Critical points at infinity and blow up of solutions of autonomous polynomial differential systems via compactification},
	journal = {Journal of Mathematical Analysis and Applications},
	volume = {318},
	number = {1},
	pages = {305-322},
	year = {2006},
	issn = {0022-247X},
	doi = {https://doi.org/10.1016/j.jmaa.2005.06.002},
	url = {https://www.sciencedirect.com/science/article/pii/S0022247X0500524X},
	author = {Uri Elias and Harry Gingold}
}

@article{BARREIRA20204416,
	title = {Bounded polynomial vector fields in R2 and Rn},
	journal = {Journal of Differential Equations},
	volume = {268},
	number = {8},
	pages = {4416-4422},
	year = {2020},
	issn = {0022-0396},
	doi = {https://doi.org/10.1016/j.jde.2019.10.023},
	url = {https://www.sciencedirect.com/science/article/pii/S0022039619305054},
	author = {Luis Barreira and Jaume Llibre and Claudia Valls}
}

@article{Stachowski_2016,
   title={Dynamical system approach to running \(\Lambda\) cosmological models},
   volume={76},
   ISSN={1434-6052},
   url={http://dx.doi.org/10.1140/epjc/s10052-016-4439-4},
   DOI={10.1140/epjc/s10052-016-4439-4},
   number={11},
   journal={The European Physical Journal C},
   publisher={Springer Science and Business Media LLC},
   author={Stachowski, Aleksander and Szydłowski, Marek},
   year={2016},
   month=nov }

@article{Berera:1995ie,
    author = "Berera, Arjun",
    title = "{Warm inflation}",
    eprint = "astro-ph/9509049",
    archivePrefix = "arXiv",
    reportNumber = "PSU-TH-159",
    doi = "10.1103/PhysRevLett.75.3218",
    journal = "Phys. Rev. Lett.",
    volume = "75",
    pages = "3218--3221",
    year = "1995"
}

@article{Kamali:2023lzq,
    author = "Kamali, Vahid and Motaharfar, Meysam and O. Ramos, Rudnei",
    title = "{Recent Developments in Warm Inflation}",
    eprint = "2302.02827",
    archivePrefix = "arXiv",
    primaryClass = "hep-ph",
    doi = "10.3390/universe9030124",
    journal = "Universe",
    volume = "9",
    number = "3",
    pages = "124",
    year = "2023"
}

@article{Mukhanov:1990me,
    author = "Mukhanov, Viatcheslav F. and Feldman, H. A. and Brandenberger, Robert H.",
    title = "{Theory of cosmological perturbations. Part 1. Classical perturbations. Part 2. Quantum theory of perturbations. Part 3. Extensions}",
    reportNumber = "BROWN-HET-796, BROWN-HET-800, BROWN-HET-780",
    doi = "10.1016/0370-1573(92)90044-Z",
    journal = "Phys. Rept.",
    volume = "215",
    pages = "203--333",
    year = "1992"
}

@article{Urena-Lopez:2007zal,
	author = "Urena-Lopez, L. Arturo and Reyes-Ibarra, Mayra J.",
	title = "{On the dynamics of a quadratic scalar field potential}",
	eprint = "0709.3996",
	archivePrefix = "arXiv",
	primaryClass = "astro-ph",
	doi = "10.1142/S0218271809014674",
	journal = "Int. J. Mod. Phys. D",
	volume = "18",
	pages = "621--634",
	year = "2009"
}

@article{Liddle_1994,
   title={Formalizing the slow-roll approximation in inflation},
   volume={50},
   ISSN={0556-2821},
   url={http://dx.doi.org/10.1103/PhysRevD.50.7222},
   DOI={10.1103/physrevd.50.7222},
   number={12},
   journal={Physical Review D},
   publisher={American Physical Society (APS)},
   author={Liddle, Andrew R. and Parsons, Paul and Barrow, John D.},
   year={1994},
   month=dec, pages={7222–7232} }

@article{Berera:2008ar,
    author = "Berera, Arjun and Moss, Ian G. and Ramos, Rudnei O.",
    title = "{Warm Inflation and its Microphysical Basis}",
    eprint = "0808.1855",
    archivePrefix = "arXiv",
    primaryClass = "hep-ph",
    doi = "10.1088/0034-4885/72/2/026901",
    journal = "Rept. Prog. Phys.",
    volume = "72",
    pages = "026901",
    year = "2009"
}

@article{Bastero-Gil:2016qru,
    author = "Bastero-Gil, Mar and Berera, Arjun and Ramos, Rudnei O. and Rosa, Joao G.",
    title = "{Warm Little Inflaton}",
    eprint = "1604.08838",
    archivePrefix = "arXiv",
    primaryClass = "hep-ph",
    doi = "10.1103/PhysRevLett.117.151301",
    journal = "Phys. Rev. Lett.",
    volume = "117",
    number = "15",
    pages = "151301",
    year = "2016"
}

@article{Berghaus:2019whh,
    author = "Berghaus, Kim V. and Graham, Peter W. and Kaplan, David E.",
    title = "{Minimal Warm Inflation}",
    eprint = "1910.07525",
    archivePrefix = "arXiv",
    primaryClass = "hep-ph",
    doi = "10.1088/1475-7516/2020/03/034",
    journal = "JCAP",
    volume = "03",
    pages = "034",
    year = "2020",
    note = "[Erratum: JCAP 10, E02 (2023)]"
}

@article{Berera:1995wh,
    author = "Berera, Arjun and Fang, Li-Zhi",
    title = "{Thermally induced density perturbations in the inflation era}",
    eprint = "astro-ph/9501024",
    archivePrefix = "arXiv",
    reportNumber = "AZPH-TH-94-8",
    doi = "10.1103/PhysRevLett.74.1912",
    journal = "Phys. Rev. Lett.",
    volume = "74",
    pages = "1912--1915",
    year = "1995"
}

@article{Berera:1996nv,
    author = "Berera, Arjun",
    title = "{Thermal properties of an inflationary universe}",
    eprint = "hep-th/9601134",
    archivePrefix = "arXiv",
    reportNumber = "PSU-TH-167",
    doi = "10.1103/PhysRevD.54.2519",
    journal = "Phys. Rev. D",
    volume = "54",
    pages = "2519--2534",
    year = "1996"
}

@article{Berera_2023,
   title={The Warm Inflation Story},
   volume={9},
   ISSN={2218-1997},
   url={http://dx.doi.org/10.3390/universe9060272},
   DOI={10.3390/universe9060272},
   number={6},
   journal={Universe},
   publisher={MDPI AG},
   author={Berera, Arjun},
   year={2023},
   month=jun, pages={272} }

@article{Bastero_Gil_2013,
   title={General dissipation coefficient in low-temperature warm inflation},
   volume={2013},
   ISSN={1475-7516},
   url={http://dx.doi.org/10.1088/1475-7516/2013/01/016},
   DOI={10.1088/1475-7516/2013/01/016},
   number={01},
   journal={Journal of Cosmology and Astroparticle Physics},
   publisher={IOP Publishing},
   author={Bastero-Gil, Mar and Berera, Arjun and Ramos, Rudnei O and Rosa, João G},
   year={2013},
   month=jan, pages={016–016} }

@article{Berera:1999ws,
	author = "Berera, Arjun",
	title = "{Warm inflation at arbitrary adiabaticity: A Model, an existence proof for inflationary dynamics in quantum field theory}",
	eprint = "hep-ph/9904409",
	archivePrefix = "arXiv",
	reportNumber = "VAND-TH-99-04",
	doi = "10.1016/S0550-3213(00)00411-9",
	journal = "Nucl. Phys. B",
	volume = "585",
	pages = "666--714",
	year = "2000"
}

@article{PhysRevD.62.083517,
  title = {Perturbation spectra in the warm inflationary scenario},
  author = {Taylor, A. N. and Berera, A.},
  journal = {Phys. Rev. D},
  volume = {62},
  issue = {8},
  pages = {083517},
  numpages = {11},
  year = {2000},
  month = {Sep},
  publisher = {American Physical Society},
  doi = {10.1103/PhysRevD.62.083517},
  url = {https://link.aps.org/doi/10.1103/PhysRevD.62.083517}
}

@article{PhysRevD.69.083525,
  title = {Scalar perturbation spectra from warm inflation},
  author = {Hall, Lisa M. H. and Moss, Ian G. and Berera, Arjun},
  journal = {Phys. Rev. D},
  volume = {69},
  issue = {8},
  pages = {083525},
  numpages = {11},
  year = {2004},
  month = {Apr},
  publisher = {American Physical Society},
  doi = {10.1103/PhysRevD.69.083525},
  url = {https://link.aps.org/doi/10.1103/PhysRevD.69.083525}
}

@article{Gupta:2002kn,
    author = "Gupta, Sujata and Berera, A. and Heavens, A. F. and Matarrese, S.",
    title = "{Non-Gaussian signatures in the cosmic background radiation from warm inflation}",
    eprint = "astro-ph/0205152",
    archivePrefix = "arXiv",
    doi = "10.1103/PhysRevD.66.043510",
    journal = "Phys. Rev. D",
    volume = "66",
    pages = "043510",
    year = "2002"
}

@article{Chen_2008,
   title={Inflationary non-Gaussianity from thermal fluctuations},
   volume={2008},
   ISSN={1475-7516},
   url={http://dx.doi.org/10.1088/1475-7516/2008/05/014},
   DOI={10.1088/1475-7516/2008/05/014},
   number={05},
   journal={Journal of Cosmology and Astroparticle Physics},
   publisher={IOP Publishing},
   author={Chen, Bin and Wang, Yi and Xue, Wei},
   year={2008},
   month=may, pages={014} }

@article{Moss:2007cv,
    author = "Moss, Ian G and Xiong, Chun",
    title = "{Non-Gaussianity in fluctuations from warm inflation}",
    eprint = "astro-ph/0701302",
    archivePrefix = "arXiv",
    doi = "10.1088/1475-7516/2007/04/007",
    journal = "JCAP",
    volume = "04",
    pages = "007",
    year = "2007"
}

@article{Moss:2011qc,
    author = "Moss, Ian G. and Yeomans, Timothy",
    title = "{Non-gaussianity in the strong regime of warm inflation}",
    eprint = "1102.2833",
    archivePrefix = "arXiv",
    primaryClass = "astro-ph.CO",
    doi = "10.1088/1475-7516/2011/08/009",
    journal = "JCAP",
    volume = "08",
    pages = "009",
    year = "2011"
}

@article{Bastero-Gil:2009sdq,
    author = "Bastero-Gil, Mar and Berera, Arjun",
    title = "{Warm inflation model building}",
    eprint = "0902.0521",
    archivePrefix = "arXiv",
    primaryClass = "hep-ph",
    doi = "10.1142/S0217751X09044206",
    journal = "Int. J. Mod. Phys. A",
    volume = "24",
    pages = "2207--2240",
    year = "2009"
}

@book{Coley:2003mj,
	author = "Coley, A. A.",
	title = "{Dynamical systems and cosmology}",
	doi = "10.1007/978-94-017-0327-7",
	publisher = "Kluwer",
	address = "Dordrecht, Netherlands",
	year = "2003"
}

@article{Rendall:2001it,
	author = "Rendall, Alan D.",
	title = "{Cosmological models and center manifold theory}",
	eprint = "gr-qc/0112040",
	archivePrefix = "arXiv",
	reportNumber = "AEI-2001-144, AEI-2001-144",
	doi = "10.1023/A:1019734703162",
	journal = "Gen. Rel. Grav.",
	volume = "34",
	pages = "1277--1294",
	year = "2002"
}

@inbook{Boehmer:2014vea,
    author = "Boehmer, Christian G. and Chan, Nyein",
    title = "{Dynamical systems in cosmology.}",
    eprint = "1409.5585",
    archivePrefix = "arXiv",
    primaryClass = "gr-qc",
    doi = "10.1142/9781786341044_0004",
    year = "2017"
}

@article{Boehmer:2011tp,
    author = "Boehmer, Christian G. and Chan, Nyein and Lazkoz, Ruth",
    title = "{Dynamics of dark energy models and centre manifolds}",
    eprint = "1111.6247",
    archivePrefix = "arXiv",
    primaryClass = "gr-qc",
    doi = "10.1016/j.physletb.2012.06.064",
    journal = "Phys. Lett. B",
    volume = "714",
    pages = "11--17",
    year = "2012"
}

@article{Bahamonde:2017ize,
    author = {Bahamonde, Sebastian and B\"ohmer, Christian G. and Carloni, Sante and Copeland, Edmund J. and Fang, Wei and Tamanini, Nicola},
    title = "{Dynamical systems applied to cosmology: dark energy and modified gravity}",
    eprint = "1712.03107",
    archivePrefix = "arXiv",
    primaryClass = "gr-qc",
    doi = "10.1016/j.physrep.2018.09.001",
    journal = "Phys. Rept.",
    volume = "775-777",
    pages = "1--122",
    year = "2018"
}

@article{Dutta:2017wfd,
	author = "Dutta, Jibitesh and Khyllep, Wompherdeiki and Tamanini, Nicola",
	title = "{Dark energy with a gradient coupling to the dark matter fluid: cosmological dynamics and structure formation}",
	eprint = "1707.09246",
	archivePrefix = "arXiv",
	primaryClass = "gr-qc",
	doi = "10.1088/1475-7516/2018/01/038",
	journal = "JCAP",
	volume = "01",
	pages = "038",
	year = "2018"
}

@article{Bouhmadi-Lopez:2016dzw,
    author = "Bouhmadi-L\'opez, Mariam and Marto, Jo\~ao and Morais, Jo\~ao and Silva, C\'esar M.",
    title = "{Cosmic infinity: A dynamical system approach}",
    eprint = "1611.03100",
    archivePrefix = "arXiv",
    primaryClass = "gr-qc",
    doi = "10.1088/1475-7516/2017/03/042",
    journal = "JCAP",
    volume = "03",
    pages = "042",
    year = "2017"
}

@book{blackmore2011nonlinear,
	title={Nonlinear dynamical systems of mathematical physics: spectral and symplectic integrability analysis},
	author={Blackmore, Denis L and Samoylenko, Valeriy Hr and others},
	year={2011},
	publisher={World Scientific}
}

@article{elias2006critical,
	title={Critical points at infinity and blow up of solutions of autonomous polynomial differential systems via compactification},
	author={Elias, Uri and Gingold, Harry},
	journal={Journal of mathematical analysis and applications},
	volume={318},
	number={1},
	pages={305--322},
	year={2006},
	publisher={Elsevier}
}

@article{Chatterjee:2021ijw,
	author = "Chatterjee, Anirban and Hussain, Saddam and Bhattacharya, Kaushik",
	title = "{Dynamical stability of the k-essence field interacting nonminimally with a perfect fluid}",
	eprint = "2105.00361",
	archivePrefix = "arXiv",
	primaryClass = "gr-qc",
	doi = "10.1103/PhysRevD.104.103505",
	journal = "Phys. Rev. D",
	volume = "104",
	number = "10",
	pages = "103505",
	year = "2021"
}

@article{Hussain:2022dhp,
	author = "Hussain, Saddam and Chakraborty, Saikat and Roy, Nandan and Bhattacharya, Kaushik",
	title = "{Dynamical systems analysis of tachyon-dark-energy models from a new perspective}",
	eprint = "2208.10352",
	archivePrefix = "arXiv",
	primaryClass = "gr-qc",
	doi = "10.1103/PhysRevD.107.063515",
	journal = "Phys. Rev. D",
	volume = "107",
	number = "6",
	pages = "063515",
	year = "2023"
}

@article{Hussain:2023kwk,
	author = "Hussain, Saddam and Chatterjee, Anirban and Bhattacharya, Kaushik",
	title = "{Dynamical stability in models where dark matter and dark energy are nonminimally coupled to curvature}",
	eprint = "2305.19062",
	archivePrefix = "arXiv",
	primaryClass = "gr-qc",
	doi = "10.1103/PhysRevD.108.103502",
	journal = "Phys. Rev. D",
	volume = "108",
	number = "10",
	pages = "103502",
	year = "2023"
}

@phdthesis{Saddam:2024xie,
	author = "Saddam, Md",
	title = "{Effects of non-minimal coupling of fluid and scalar field in cosmology}",
	school = "Indian Institute of Technology Kanpur",
	month = "6",
	year = "2024",
    url = {http://hdl.handle.net/10603/668204}
}

@article{Hussain:2024qrd,
	author = "Hussain, Saddam and Nelleri, Sarath and Bhattacharya, Kaushik",
	title = "{Comprehensive study of k-essence model: dynamical system analysis and observational constraints from latest Type Ia supernova and BAO observations}",
	eprint = "2406.07179",
	archivePrefix = "arXiv",
	primaryClass = "astro-ph.CO",
	doi = "10.1088/1475-7516/2025/03/025",
	journal = "JCAP",
	volume = "03",
	pages = "025",
	year = "2025"
}

@article{Bhattacharya:2022wzu,
	author = "Bhattacharya, Kaushik and Chatterjee, Anirban and Hussain, Saddam",
	title = "{Dynamical stability in presence of non-minimal derivative dependent coupling of k-essence field with a relativistic fluid}",
	eprint = "2206.12398",
	archivePrefix = "arXiv",
	primaryClass = "gr-qc",
	doi = "10.1140/epjc/s10052-023-11666-w",
	journal = "Eur. Phys. J. C",
	volume = "83",
	number = "6",
	pages = "488",
	year = "2023"
}

@article{Copeland:2006wr,
	author = "Copeland, Edmund J. and Sami, M. and Tsujikawa, Shinji",
	title = "{Dynamics of dark energy}",
	eprint = "hep-th/0603057",
	archivePrefix = "arXiv",
	doi = "10.1142/S021827180600942X",
	journal = "Int. J. Mod. Phys. D",
	volume = "15",
	pages = "1753--1936",
	year = "2006"
}

@article{Motohashi_2015,
   title={Inflation with a constant rate of roll},
   volume={2015},
   ISSN={1475-7516},
   url={http://dx.doi.org/10.1088/1475-7516/2015/09/018},
   DOI={10.1088/1475-7516/2015/09/018},
   number={09},
   journal={Journal of Cosmology and Astroparticle Physics},
   publisher={IOP Publishing},
   author={Motohashi, Hayato and Starobinsky, Alexei A. and Yokoyama, Jun’ichi},
   year={2015},
   month=sep, pages={018–018} }

@article{Yi:2017mxs,
    author = "Yi, Zhu and Gong, Yungui",
    title = "{On the constant-roll inflation}",
    eprint = "1712.07478",
    archivePrefix = "arXiv",
    primaryClass = "gr-qc",
    doi = "10.1088/1475-7516/2018/03/052",
    journal = "JCAP",
    volume = "03",
    pages = "052",
    year = "2018"
}

@article{Guerrero:2020lng,
    author = "Guerrero, Merce and Rubiera-Garcia, Diego and Saez-Chillon Gomez, Diego",
    title = "{Constant roll inflation in multifield models}",
    eprint = "2008.07260",
    archivePrefix = "arXiv",
    primaryClass = "gr-qc",
    doi = "10.1103/PhysRevD.102.123528",
    journal = "Phys. Rev. D",
    volume = "102",
    pages = "123528",
    year = "2020"
}

@article{Motohashi_2017,
   title={f(R) constant-roll inflation},
   volume={77},
   ISSN={1434-6052},
   url={http://dx.doi.org/10.1140/epjc/s10052-017-5109-x},
   DOI={10.1140/epjc/s10052-017-5109-x},
   number={8},
   journal={The European Physical Journal C},
   publisher={Springer Science and Business Media LLC},
   author={Motohashi, Hayato and Starobinsky, Alexei A.},
   year={2017},
   month=aug }

@article{Motohashi_20171,
   title={Constant-roll inflation: Confrontation with recent observational data},
   volume={117},
   ISSN={1286-4854},
   url={http://dx.doi.org/10.1209/0295-5075/117/39001},
   DOI={10.1209/0295-5075/117/39001},
   number={3},
   journal={EPL (Europhysics Letters)},
   publisher={IOP Publishing},
   author={Motohashi, Hayato and Starobinsky, Alexei A.},
   year={2017},
   month=feb, pages={39001} }

@article{Oikonomou_2017,
   title={Reheating in constant-roll F(R) gravity},
   volume={32},
   ISSN={1793-6632},
   url={http://dx.doi.org/10.1142/S0217732317501723},
   DOI={10.1142/s0217732317501723},
   number={33},
   journal={Modern Physics Letters A},
   publisher={World Scientific Pub Co Pte Ltd},
   author={Oikonomou, V. K.},
   year={2017},
   month=oct, pages={1750172} }

@article{Gao_2017,
   title={Reconstruction of constant slow-roll inflation},
   volume={60},
   ISSN={1869-1927},
   url={http://dx.doi.org/10.1007/s11433-017-9065-4},
   DOI={10.1007/s11433-017-9065-4},
   number={9},
   journal={Science China Physics, Mechanics \&amp; Astronomy},
   publisher={Springer Science and Business Media LLC},
   author={Gao, Qing},
   year={2017},
   month=jul }

@article{Odintsov_2017,
   title={Inflation with a smooth constant-roll to constant-roll era transition},
   volume={96},
   ISSN={2470-0029},
   url={http://dx.doi.org/10.1103/PhysRevD.96.024029},
   DOI={10.1103/physrevd.96.024029},
   number={2},
   journal={Physical Review D},
   publisher={American Physical Society (APS)},
   author={Odintsov, S. D. and Oikonomou, V. K.},
   year={2017},
   month=jul }

@article{Ito_2018,
   title={Anisotropic constant-roll Inflation},
   volume={78},
   ISSN={1434-6052},
   url={http://dx.doi.org/10.1140/epjc/s10052-018-5534-5},
   DOI={10.1140/epjc/s10052-018-5534-5},
   number={1},
   journal={The European Physical Journal C},
   publisher={Springer Science and Business Media LLC},
   author={Ito, Asuka and Soda, Jiro},
   year={2018},
   month=jan }

@article{Karam_2018,
   title={Constant-roll (quasi-)linear inflation},
   volume={2018},
   ISSN={1475-7516},
   url={http://dx.doi.org/10.1088/1475-7516/2018/05/011},
   DOI={10.1088/1475-7516/2018/05/011},
   number={05},
   journal={Journal of Cosmology and Astroparticle Physics},
   publisher={IOP Publishing},
   author={Karam, A. and Marzola, L. and Pappas, T. and Racioppi, A. and Tamvakis, K.},
   year={2018},
   month=may, pages={011–011} }

@article{Cicciarella_2018,
   title={New perspectives on constant-roll inflation},
   volume={2018},
   ISSN={1475-7516},
   url={http://dx.doi.org/10.1088/1475-7516/2018/01/024},
   DOI={10.1088/1475-7516/2018/01/024},
   number={01},
   journal={Journal of Cosmology and Astroparticle Physics},
   publisher={IOP Publishing},
   author={Cicciarella, Francesco and Mabillard, Joel and Pieroni, Mauro},
   year={2018},
   month=jan, pages={024–024} }

@article{Anguelova_2018,
   title={Systematics of constant roll inflation},
   volume={2018},
   ISSN={1475-7516},
   url={http://dx.doi.org/10.1088/1475-7516/2018/02/004},
   DOI={10.1088/1475-7516/2018/02/004},
   number={02},
   journal={Journal of Cosmology and Astroparticle Physics},
   publisher={IOP Publishing},
   author={Anguelova, Lilia and Suranyi, Peter and Wijewardhana, L.C.R.},
   year={2018},
   month=feb, pages={004–004} }

@article{Gao_2018,
   title={The observational constraint on constant-roll inflation},
   volume={61},
   ISSN={1869-1927},
   url={http://dx.doi.org/10.1007/s11433-018-9197-2},
   DOI={10.1007/s11433-018-9197-2},
   number={7},
   journal={Science China Physics, Mechanics \&amp; Astronomy},
   publisher={Springer Science and Business Media LLC},
   author={Gao, Qing},
   year={2018},
   month=mar }

@article{Gao_2018_1,
   title={The observational constraint on constant-roll inflation},
   volume={61},
   ISSN={1869-1927},
   url={http://dx.doi.org/10.1007/s11433-018-9197-2},
   DOI={10.1007/s11433-018-9197-2},
   number={7},
   journal={Science China Physics, Mechanics \&amp; Astronomy},
   publisher={Springer Science and Business Media LLC},
   author={Gao, Qing},
   year={2018},
   month=mar }

@article{Awad_2018,
   title={Constant-roll inflation in f(T) teleparallel gravity},
   volume={2018},
   ISSN={1475-7516},
   url={http://dx.doi.org/10.1088/1475-7516/2018/07/026},
   DOI={10.1088/1475-7516/2018/07/026},
   number={07},
   journal={Journal of Cosmology and Astroparticle Physics},
   publisher={IOP Publishing},
   author={Awad, A. and Hanafy, W. El and Nashed, G.G.L. and Odintsov, S.D. and Oikonomou, V.K.},
   year={2018},
   month=jul, pages={026–026} }

@article{Odintsov_2017_02,
   title={Unification of constant-roll inflation and dark energy with logarithmic R2-corrected and exponential F(R) gravity},
   volume={923},
   ISSN={0550-3213},
   url={http://dx.doi.org/10.1016/j.nuclphysb.2017.08.018},
   DOI={10.1016/j.nuclphysb.2017.08.018},
   journal={Nuclear Physics B},
   publisher={Elsevier BV},
   author={Odintsov, S.D. and Oikonomou, V.K. and Sebastiani, L.},
   year={2017},
   month=oct, pages={608–632} }

@article{Nojiri_2017,
   title={Constant-roll inflation in F(
                    R
                    ) gravity},
   volume={34},
   ISSN={1361-6382},
   url={http://dx.doi.org/10.1088/1361-6382/aa92a4},
   DOI={10.1088/1361-6382/aa92a4},
   number={24},
   journal={Classical and Quantum Gravity},
   publisher={IOP Publishing},
   author={Nojiri, S and Odintsov, S D and Oikonomou, V K},
   year={2017},
   month=nov, pages={245012} }

@article{Mohammadi:2022tmk,
    author = "Mohammadi, Ali and Ahmadi, Nahid and Shokri, Mehdi",
    title = "{On the constant roll complex scalar field inflationary models}",
    eprint = "2212.13403",
    archivePrefix = "arXiv",
    primaryClass = "gr-qc",
    doi = "10.1088/1475-7516/2023/06/058",
    journal = "JCAP",
    volume = "06",
    pages = "058",
    year = "2023"
}

@article{Odintsov_2017_01,
   title={Autonomous dynamical system approach for 
 gravity},
   volume={96},
   ISSN={2470-0029},
   url={http://dx.doi.org/10.1103/PhysRevD.96.104049},
   DOI={10.1103/physrevd.96.104049},
   number={10},
   journal={Physical Review D},
   publisher={American Physical Society (APS)},
   author={Odintsov, S. D. and Oikonomou, V. K.},
   year={2017},
   month=nov }

@article{Shokri:2021zqw,
    author = "Shokri, Mehdi and Sadeghi, Jafar and Gashti, Saeed Noori",
    title = "{Quintessential constant-roll inflation}",
    eprint = "2107.04756",
    archivePrefix = "arXiv",
    primaryClass = "astro-ph.CO",
    doi = "10.1016/j.dark.2021.100923",
    journal = "Phys. Dark Univ.",
    volume = "35",
    pages = "100923",
    year = "2022"
}

@article{Lin:2019fcz,
    author = "Lin, Wei-Chen and Morse, Michael J. P. and Kinney, William H.",
    title = "{Dynamical Analysis of Attractor Behavior in Constant Roll Inflation}",
    eprint = "1904.06289",
    archivePrefix = "arXiv",
    primaryClass = "astro-ph.CO",
    doi = "10.1088/1475-7516/2019/09/063",
    journal = "JCAP",
    volume = "09",
    pages = "063",
    year = "2019"
}

@article{Morse:2018kda,
    author = "Morse, Michael J. P. and Kinney, William H.",
    title = "{Large-$\eta$ constant-roll inflation is never an attractor}",
    eprint = "1804.01927",
    archivePrefix = "arXiv",
    primaryClass = "astro-ph.CO",
    doi = "10.1103/PhysRevD.97.123519",
    journal = "Phys. Rev. D",
    volume = "97",
    number = "12",
    pages = "123519",
    year = "2018"
}

@article{deOliveira:1997jt,
    author = "de Oliveira, H. P. and Ramos, Rudnei O.",
    title = "{Dynamical system analysis for inflation with dissipation}",
    eprint = "gr-qc/9710093",
    archivePrefix = "arXiv",
    reportNumber = "IF-UERJ-97-12",
    doi = "10.1103/PhysRevD.57.741",
    journal = "Phys. Rev. D",
    volume = "57",
    pages = "741--749",
    year = "1998"
}

@article{Moss:2008yb,
    author = "Moss, Ian G. and Xiong, Chun",
    title = "{On the consistency of warm inflation}",
    eprint = "0808.0261",
    archivePrefix = "arXiv",
    primaryClass = "astro-ph",
    doi = "10.1088/1475-7516/2008/11/023",
    journal = "JCAP",
    volume = "11",
    pages = "023",
    year = "2008"
}

@article{Li:2018sfs,
	author = "Li, Xi-Bin and Wang, Yang-Yang and Wang, He and Zhu, Jian-Yang",
	title = "{Dynamic analysis of noncanonical warm inflation}",
	eprint = "1804.05360",
	archivePrefix = "arXiv",
	primaryClass = "gr-qc",
	doi = "10.1103/PhysRevD.98.043510",
	journal = "Phys. Rev. D",
	volume = "98",
	number = "4",
	pages = "043510",
	year = "2018"
}

@article{Das:2023rat,
	author = "Das, Suratna and Hussain, Saddam and Nandi, Debottam and O. Ramos, Rudnei and Silva, Renato",
	title = "{Stability analysis of warm quintessential dark energy model}",
	eprint = "2306.09369",
	archivePrefix = "arXiv",
	primaryClass = "gr-qc",
	doi = "10.1103/PhysRevD.108.083517",
	journal = "Phys. Rev. D",
	volume = "108",
	number = "8",
	pages = "083517",
	year = "2023"
}

@article{Odintsov:2023lbb,
	author = "Odintsov, Sergei D. and Paul, Tanmoy",
	title = "{From inflation to reheating and their dynamical stability analysis in Gauss\textendash{}Bonnet gravity}",
	eprint = "2305.19110",
	archivePrefix = "arXiv",
	primaryClass = "gr-qc",
	doi = "10.1016/j.dark.2023.101263",
	journal = "Phys. Dark Univ.",
	volume = "42",
	pages = "101263",
	year = "2023"
}

@article{DAgostino:2021vvv,
	author = "D'Agostino, Rocco and Luongo, Orlando",
	title = "{Cosmological viability of a double field unified model from warm inflation}",
	eprint = "2112.12816",
	archivePrefix = "arXiv",
	primaryClass = "astro-ph.CO",
	doi = "10.1016/j.physletb.2022.137070",
	journal = "Phys. Lett. B",
	volume = "829",
	pages = "137070",
	year = "2022"
}

@article{Zhang:2024kcf,
	author = "Zhang, Xiao-Min and Zhao, Run-Qing and Peng, Zhi-peng and Li, Xi-Bin and Feng, Yun-Cai and Chu, Peng-Cheng and Xing, Yi-Hang",
	title = "{Noncanonical warm inflation with nonminimal derivative coupling}",
	eprint = "2410.16839",
	archivePrefix = "arXiv",
	primaryClass = "gr-qc",
	doi = "10.1088/1475-7516/2025/02/013",
	journal = "JCAP",
	volume = "02",
	pages = "013",
	year = "2025"
}

@article{Jawad:2017nkq,
	author = "Jawad, Abdul and Chaudhary, Shahid and Videla, Nelson",
	title = "{Dynamics of polynomial chaplygin gas warm inflation}",
	eprint = "1711.03879",
	archivePrefix = "arXiv",
	primaryClass = "gr-qc",
	doi = "10.1140/epjc/s10052-017-5377-5",
	journal = "Eur. Phys. J. C",
	volume = "77",
	number = "11",
	pages = "808",
	year = "2017"
}

@article{Kamali:2019wdh,
    author = "Kamali, Vahid and Artymowski, Micha\l{} and Setare, Mohammad Reza",
    title = "{Constant roll warm inflation in high dissipative regime}",
    eprint = "1905.04814",
    archivePrefix = "arXiv",
    primaryClass = "gr-qc",
    reportNumber = "JCAP07(2020)002",
    doi = "10.1088/1475-7516/2020/07/002",
    journal = "JCAP",
    volume = "07",
    number = "07",
    pages = "002",
    year = "2020"
}

@article{Kofman:1997yn,
    author = "Kofman, Lev and Linde, Andrei D. and Starobinsky, Alexei A.",
    title = "{Towards the theory of reheating after inflation}",
    eprint = "hep-ph/9704452",
    archivePrefix = "arXiv",
    reportNumber = "IFA-97-28, SU-ITP-97-18",
    doi = "10.1103/PhysRevD.56.3258",
    journal = "Phys. Rev. D",
    volume = "56",
    pages = "3258--3295",
    year = "1997"
}

@article{Kofman:1994rk,
    author = "Kofman, Lev and Linde, Andrei D. and Starobinsky, Alexei A.",
    title = "{Reheating after inflation}",
    eprint = "hep-th/9405187",
    archivePrefix = "arXiv",
    reportNumber = "UH-IFA-94-35, SU-ITP-94-13, YITP-U-94-15",
    doi = "10.1103/PhysRevLett.73.3195",
    journal = "Phys. Rev. Lett.",
    volume = "73",
    pages = "3195--3198",
    year = "1994"
}

@article{Allahverdi:2010xz,
    author = "Allahverdi, Rouzbeh and Brandenberger, Robert and Cyr-Racine, Francis-Yan and Mazumdar, Anupam",
    title = "{Reheating in Inflationary Cosmology: Theory and Applications}",
    eprint = "1001.2600",
    archivePrefix = "arXiv",
    primaryClass = "hep-th",
    doi = "10.1146/annurev.nucl.012809.104511",
    journal = "Ann. Rev. Nucl. Part. Sci.",
    volume = "60",
    pages = "27--51",
    year = "2010"
}

@article{Bassett:2005xm,
    author = "Bassett, Bruce A. and Tsujikawa, Shinji and Wands, David",
    title = "{Inflation dynamics and reheating}",
    eprint = "astro-ph/0507632",
    archivePrefix = "arXiv",
    doi = "10.1103/RevModPhys.78.537",
    journal = "Rev. Mod. Phys.",
    volume = "78",
    pages = "537--589",
    year = "2006"
}

@article{Biswas_2024,
   title={Embedding ultraslow-roll inflaton dynamics in warm inflation},
   volume={109},
   ISSN={2470-0029},
   url={http://dx.doi.org/10.1103/PhysRevD.109.023501},
   DOI={10.1103/physrevd.109.023501},
   number={2},
   journal={Physical Review D},
   publisher={American Physical Society (APS)},
   author={Biswas, Sandip and Bhattacharya, Kaushik and Das, Suratna},
   year={2024},
   month=jan }

@article{Pattison_2018,
   title={The attractive behaviour of ultra-slow-roll inflation},
   volume={2018},
   ISSN={1475-7516},
   url={http://dx.doi.org/10.1088/1475-7516/2018/08/048},
   DOI={10.1088/1475-7516/2018/08/048},
   number={08},
   journal={Journal of Cosmology and Astroparticle Physics},
   publisher={IOP Publishing},
   author={Pattison, Chris and Vennin, Vincent and Assadullahi, Hooshyar and Wands, David},
   year={2018},
   month=aug, pages={048–048} }

@article{Dimopoulos_2017,
   title={Ultra slow-roll inflation demystified},
   volume={775},
   ISSN={0370-2693},
   url={http://dx.doi.org/10.1016/j.physletb.2017.10.066},
   DOI={10.1016/j.physletb.2017.10.066},
   journal={Physics Letters B},
   publisher={Elsevier BV},
   author={Dimopoulos, Konstantinos},
   year={2017},
   month=dec, pages={262–265} }

@article{Mishra_2020,
   title={Primordial black holes from a tiny bump/dip in the inflaton potential},
   volume={2020},
   ISSN={1475-7516},
   url={http://dx.doi.org/10.1088/1475-7516/2020/04/007},
   DOI={10.1088/1475-7516/2020/04/007},
   number={04},
   journal={Journal of Cosmology and Astroparticle Physics},
   publisher={IOP Publishing},
   author={Mishra, Swagat S. and Sahni, Varun},
   year={2020},
   month=apr, pages={007–007} }

@article{Passaglia:2018ixg,
    author = "Passaglia, Samuel and Hu, Wayne and Motohashi, Hayato",
    title = "{Primordial black holes and local non-Gaussianity in canonical inflation}",
    eprint = "1812.08243",
    archivePrefix = "arXiv",
    primaryClass = "astro-ph.CO",
    reportNumber = "YITP-18-128",
    doi = "10.1103/PhysRevD.99.043536",
    journal = "Phys. Rev. D",
    volume = "99",
    number = "4",
    pages = "043536",
    year = "2019"
}

@article{Cai:2018dkf,
    author = "Cai, Yi-Fu and Chen, Xingang and Namjoo, Mohammad Hossein and Sasaki, Misao and Wang, Dong-Gang and Wang, Ziwei",
    title = "{Revisiting non-Gaussianity from non-attractor inflation models}",
    eprint = "1712.09998",
    archivePrefix = "arXiv",
    primaryClass = "astro-ph.CO",
    reportNumber = "MIT-CTP-4974, YITP-17-133",
    doi = "10.1088/1475-7516/2018/05/012",
    journal = "JCAP",
    volume = "05",
    pages = "012",
    year = "2018"
}

@inproceedings{Motohashi:2025qgd,
    author = "Motohashi, Hayato",
    title = "{Constant-Roll Inflation}",
    eprint = "2504.16757",
    archivePrefix = "arXiv",
    primaryClass = "astro-ph.CO",
    month = "4",
    year = "2025",
    url={https://arxiv.org/abs/2504.16757}
}

\end{document}